\documentclass[fleqn,onecolumn]{pasj00}

\def\commenta{$^*$}
\def\commentb{$^\dagger$}
\def\commentc{$^\ddagger$}
\def\commentd{$^\S$}
\def\commente{$^\|$}
\def\commentf{$^\#$}

\def\inpress{in press}

\def\arxiv#1{ (arxiv/#1)}

\DeclareAbbreviation\AAHam{Astron. Abh. Hamburg. Sternw.}
\DeclareAbbreviation\AARv{Astron. Astrophys. Rev.}
\DeclareAbbreviation\an{Astron. Nachr.}
\DeclareAbbreviation\AcA{Acta Astron.}
\DeclareAbbreviation\actaa{Acta Astron.}
\DeclareAbbreviation\Afz{Astrofizika}
\DeclareAbbreviation\AnTok{Tokyo Astron. Obs. Annals, Sec. Ser.}
\DeclareAbbreviation\Ap{Astrophysics}
\DeclareAbbreviation\ARep{Astron. Rep.}
\DeclareAbbreviation\ATel{Astronomer's Telegram}
\DeclareAbbreviation\ATsir{Astron. Tsirk.}
\DeclareAbbreviation\AcApS{Acta Astrophys. Sinica}
\DeclareAbbreviation\AstL{Astron. Lett.}
\DeclareAbbreviation\BaltA{Baltic Astron.}
\DeclareAbbreviation\BASI{Bull. Astron. Soc. India}
\DeclareAbbreviation\BeSN{Be Newslett.}
\DeclareAbbreviation\caa{Chinese J. Astron. Astrophys.}
\DeclareAbbreviation\GCN{GRB Coord. Netw. Circ.}
\DeclareAbbreviation\ibvs{Inf. Bull. Variable Stars}
\DeclareAbbreviation\JAD{J. Astron. Data}
\DeclareAbbreviation\JAVSO{J. American Assoc. Variable Star Obs.}
\DeclareAbbreviation\JBAA{J. British Astron. Assoc.}
\DeclareAbbreviation\JO{Journal des Observateurs}
\DeclareAbbreviation\LowOB{Lowell Obs. Bull.}
\DeclareAbbreviation\MitVS{Mitteil. Ver\"{a}nderl. Sterne}
\DeclareAbbreviation\MmSAI{Mem. Soc. Astron. Ita.}
\DeclareAbbreviation\Msngr{Messenger}
\DeclareAbbreviation\NewA{New Astron.}
\DeclareAbbreviation\na{New Astron.}
\DeclareAbbreviation\NewAR{New Astron. Rev.}
\DeclareAbbreviation\NZJS{New Zealand Journal of Science}
\DeclareAbbreviation\OAP{Odessa Astron. Publ.}
\DeclareAbbreviation\Obs{Observatory}
\DeclareAbbreviation\PASA{Publ. Astron. Soc. Australia}
\DeclareAbbreviation\PAZh{Pis'ma AZh}
\DeclareAbbreviation\POBeo{Publ. Astron. Obs. Belgr.}
\DeclareAbbreviation\PhR{Phys. Rep.}
\DeclareAbbreviation\PVSS{Publ. Variable Stars Sect. R. Astron. Soc. New Zealand}
\DeclareAbbreviation\PZ{Perem. Zvezdy}
\DeclareAbbreviation\PZP{Perem. Zvezdy Pril.}
\DeclareAbbreviation\QJRAS{QJRAS}
\DeclareAbbreviation\RMxAA{Rev. Mexicana Astron. Astrof.}
\DeclareAbbreviation\RvMA{Reviews of Modern Astron.}
\DeclareAbbreviation\Sci{Science}
\DeclareAbbreviation\SvA{Soviet Astronomy}
\DeclareAbbreviation\SvAL{Soviet Astronomy Letters}
\DeclareAbbreviation\VeSon{Ver\"{o}ff. Sternw. Sonneberg}
\DeclareAbbreviation\VSOLJBul{VSOLJ Variable Star Bull.}
\DeclareAbbreviation\yCat{VizieR Online Data Catalog}
\DeclareAbbreviation\ZA{Z. Astrophys.}

\def\ASPConf#1#2{ASP Conf. Ser. #1, #2}

\def\PublisherCambridge{Cambridge: Cambridge University Press}

\def\PublisherASP{San Francisco: ASP}

\def\PublisherSpringer{Berlin: Springer-Verlag}
\def\PublisherUAP{Tokyo: Universal Academy Press}

\newcounter{author}
\def\authorcount#1#2{\refstepcounter{author}\label{#1}
                     \altaffiltext{\ref{#1}}{#2}}

\begin{document}
\SetRunningHead{Ohshima et al.}{ER UMa observations}

\Received{201X/XX/XX}
\Accepted{201X/XX/XX}
\title{Study of Negative and Positive Superhumps in ER Ursae Majoris}

\author{Tomohito~\textsc{Ohshima},\altaffilmark{\ref{affil:Kyoto}*}
        Taichi~\textsc{Kato},\altaffilmark{\ref{affil:Kyoto}}
        Elena~\textsc{Pavlenko},\altaffilmark{\ref{affil:CrAO}}
         Hidehiko~\textsc{Akazawa},\altaffilmark{\ref{affil:OUS}}
         Kazuyoshi~\textsc{Imamura},\altaffilmark{\ref{affil:OUS}}
         Kenji~\textsc{Tanabe},\altaffilmark{\ref{affil:OUS}}
         Enrique~de~\textsc{Miguel},\altaffilmark{\ref{affil:Miguel}}$^,$\altaffilmark{\ref{affil:Miguel2}}
          William~\textsc{Stein},\altaffilmark{\ref{affil:AAVSO}}
          Hiroshi~\textsc{Itoh},\altaffilmark{\ref{affil:Ioh}}
         Franz-Josef~\textsc{Hambsch},\altaffilmark{\ref{affil:AAVSO}}$^,$\altaffilmark{\ref{affil:Hambsch}}
         Pavol~A.~\textsc{Dubovsky},\altaffilmark{\ref{affil:DPV}}
         Igor~\textsc{Kudzej}, \altaffilmark{\ref{affil:DPV}}
        Thomas~\textsc{Krajci},\altaffilmark{\ref{affil:Kra}}
       Alex~\textsc{Baklanov},\altaffilmark{\ref{affil:CrAO}}
         Denis~\textsc{Samsonov},\altaffilmark{\ref{affil:CrAO}}
         Oksana~\textsc{Antonyuk},\altaffilmark{\ref{affil:CrAO}}
         Viktor~\textsc{Malanushenko},\altaffilmark{\ref{affil:CrAO}}
         Maksim~\textsc{Andreev},\altaffilmark{\ref{affil:Terskol}}
         Ryo~\textsc{ Noguchi},\altaffilmark{\ref{affil:OKU}}
         Kazuyuki~\textsc{Ogura},\altaffilmark{\ref{affil:OKU}}
         Takashi~\textsc{Nomoto},\altaffilmark{\ref{affil:OKU}}
         Rikako~\textsc{Ono},\altaffilmark{\ref{affil:OKU}}
         Shin'ichi~\textsc{Nakagawa},\altaffilmark{\ref{affil:OKU}}
         Keisuke~\textsc{Taniuchi},\altaffilmark{\ref{affil:OKU}}
         Tomoya~\textsc{Aoki},\altaffilmark{\ref{affil:OKU}}
         Miho~\textsc{Kawabata},\altaffilmark{\ref{affil:OKU}}
         Hitoshi~\textsc{Kimura},\altaffilmark{\ref{affil:OKU}}
         Kazunari~\textsc{Masumoto},\altaffilmark{\ref{affil:OKU}}
         Hiroshi~\textsc{ Kobayashi},\altaffilmark{\ref{affil:OKU}}
         Katsura~\textsc{Matsumoto},\altaffilmark{\ref{affil:OKU}}
         Kazuhiko~\textsc{Shiokawa},\altaffilmark{\ref{affil:VSOLJ}} 
         Sergey~Yu.~\textsc{Shugarov},\altaffilmark{\ref{affil:Sternberg}}$^,$\altaffilmark{\ref{affil:Slovak}}
          Natalia~\textsc{Katysheva},\altaffilmark{\ref{affil:Sternberg}}
         Irina~\textsc{Voloshina},\altaffilmark{\ref{affil:Sternberg}}
         Polina~\textsc{Zemko},\altaffilmark{\ref{affil:Sternberg}}
         Kiyoshi~\textsc{Kasai},\altaffilmark{\ref{affil:VSOLJ}}
         Javier~\textsc{Ruiz},\altaffilmark{\ref{affil:Ruiz}}$^,$\altaffilmark{\ref{affil:Ruiz2}}
         Hiroyuki~\textsc{Maehara},\altaffilmark{\ref{affil:Kiso}}
         Natalia~\textsc{Virnina},\altaffilmark{\ref{affil:VIR}}
         Jani~\textsc{Virtanen},\altaffilmark{\ref{affil:Vir}}
        Ian~\textsc{Miller},\altaffilmark{\ref{affil:Miller}}
         Boyd~\textsc{Boitnott},\altaffilmark{\ref{affil:AAVSO}}
          Colin~\textsc{Littlefield},\altaffilmark{\ref{affil:LCO}}
         Nick~\textsc{James},\altaffilmark{\ref{affil:NDJ}}
        Tamas ~\textsc{Tordai},\altaffilmark{\ref{affil:Pol}}
         Fidrich~\textsc{Robaert},\altaffilmark{\ref{affil:Pol}}
         Stefono~\textsc{Padovan},\altaffilmark{\ref{affil:PSD}}
         Atsushi~\textsc{Miyashita},\altaffilmark{\ref{affil:Sac}}
 }
\authorcount{affil:Kyoto}{
     Department of Astronomy, Kyoto University, Kyoto 606-8502}
\email{$^*$ohshima@kusastro.kyoto-u.ac.jp}

\authorcount{affil:CrAO}{
     Crimean Astrophysical Observatory, 98409, Nauchny, Crimea, Ukraine}

\authorcount{affil:OUS}{
  Department of Biosphere-Geosphere System Science, Faculty of Informatics, Okayama University of Science, 1-1 Ridai-cho, Okayama, Okayama 700-0005, Japan}

\authorcount{affil:Miguel}{
     Departamento de F\'isica Aplicada, Facultad de Ciencias
     Experimentales, Universidad de Huelva,
     21071 Huelva, Spain}

\authorcount{affil:Miguel2}{
     Center for Backyard Astrophysics, Observatorio del CIECEM,
     Parque Dunar, Matalasca\~nas, 21760 Almonte, Huelva, Spain}

\authorcount{affil:AAVSO}{
American Association of Variable Star Observers (AAVSO)}

\authorcount{affil:Ioh}{
     VSOLJ, 1001-105 Nishiterakata, Hachioji, Tokyo 192-0153}

\authorcount{affil:Hambsch}{
Vereniging Voor Sterrnkunde (VVS), Oude Bleken 12, 2400 Mol, Belgium}

\authorcount{affil:DPV}{
     Vihorlat Observatory, Mierova 4, Humenne, Slovakia}

\authorcount{affil:Kra}{
  Center for Backyard Astrophysics (New Mexico), PO Box 1351, Cloudcroft, NM 83117, USA}

\authorcount{affil:Terskol}{
Institute  of Astronomy, Russian Academy of Sciences, 361605 Peak Terskol, Kabardino-Balkaria, Russia}

\authorcount{affil:OKU}{
     Osaka Kyoiku University, 4-698-1 Asahigaoka, Osaka 582-8582}

\authorcount{affil:VSOLJ}{
Variable Star Observer's League in Japan (VSOLJ)}

\authorcount{affil:Sternberg}{
     Sternberg Astronomical Institute, Lomonosov Moscow University, 
     Universitetsky Ave., 13, Moscow 119992, Russia}

\authorcount{affil:Slovak}{
     Astronomical Institute of the Slovak Academy of Sciences, 05960,
     Tatranska Lomnica, the Slovak Republic}

\authorcount{affil:Ruiz}{Observatorio de Cantabria, Ctra. de Rocamundo s/n, Valderredible, Cantabria, Spain}

\authorcount{affil:Ruiz2}{Agrupacion Astronomica Cantabra, Apartado 573, 39080-Santander, Spain.}

\authorcount{affil:Kiso}{
Kiso Observatory, Institute of Astronomy, School of Science, The University of Tokyo, 10762-30 Mitake, Kiso-machi, Kiso-gun, Nagano 397-0101
}

\authorcount{affil:VIR}{
Department of High and Applied Mathematics, Odessa National Maritime University, Ukraine}

\authorcount{affil:Vir}{Ollilantie 98, 84880 Ylivieska, Finland}

\authorcount{affil:Miller}{
     Furzehill House, Ilston, Swansea, SA2 7LE, UK}

\authorcount{affil:LCO}{
     Department of Physics, University of Notre Dame, Notre Dame,
     Indiana 46556, USA}

\authorcount{affil:NDJ}{
  1 Tavistock Road, Chelmsford, Essex CM1 6JL, UK}

\authorcount{affil:Pol}{
Polaris Observatory, Hungarian Astronomical Association, Laborc utca 2/c, 1037 Budapest, Hungary}

\authorcount{affil:PSD}{
American Association of Variable Star Observers, 49 Bay State Rd., Cambridge, MA 02138, USA}

\authorcount{affil:Sac}{
Seikei Meteorological Observatory, Seikei High School, 3-3-1, Kichijoji-Kitamachi, Musashino-shi, Tokyo 180-8633}



\maketitle

\begin{abstract}
We carried out the photometric observations of the SU UMa-type 
dwarf nova ER UMa during 2011 and 2012, which showed the existence of 
persistent negative 
superhumps even during the superoutburst. We performed
two-dimensional period analysis of its light curves by using a method
called ``least absolute shrinkage 
and selection operator'' (Lasso) and ``phase dispersion minimization''
(PDM) analysis, and we found that the period of negative superhumps 
systematically changed between a superoutburst and 
the next superoutburst.  The trend of the period change can be
interpreted as reflecting the change of the disk radius. 
This change of the disk radius is in good agreement with 
the predicted change of the disk radius by the
thermal-tidal instability (TTI) model.  The normal outbursts 
within a supercycle showed a general trend that the rising rate 
to maximum becomes slower as 
the next superoutburst approaches.  The change can be interpreted 
as the consequence of the increased gas-stream flow onto 
the inner region of the disk as the result of the tilted disk. 
Some of the superoutbursts were found to be triggered by a precursor
normal outburst when the positive superhumps appeared to
develop.  The 
positive and
negative superhumps co-existed during the superoutburst. The positive
superhumps were prominent only during four or five days after the
supermaximum, while the signal of 
the negative superhumps became strong after the middle
phase of the superoutburst plateau.  A simple combination of
the positive and negative superhumps was found to be insufficient
in reproducing the complex profile variation.
We were able to detect the developing phase of positive superhumps
(stage A superhumps) for the first time in ER UMa-type 
dwarf novae.  Using the period of stage A superhumps,
 we obtained a mass ratio
of 0.100(15), which indicates that ER UMa is on the ordinary
evolutional track of cataclysmic variable stars. 
\end{abstract}
q

\section{Introduction}

 Dwarf novae (DNe) are a class of cataclysmic variables (CVs), 
which consist of a white dwarf primary and a late-type secondary which 
fills its Roche lobe. 
The material transfered toward the primary through the inner 
Lagrangian point (L1) forms an accretion disk around     
the white-dwarf. The accretion disk causes instabilities, 
which are observed as an outburst [for reviews, see \citet{war95book};     
\citet{osa96review}; \citet{hel01book}].
  SU UMa-type stars are a subgroup of dwarf novae. They 
 are characterized by the presence of two 
types of outbursts, normal outburst and superoutburst. 
Whereas a normal outburst lasts for only a few days, a 
superoutburst lasts for about two weeks and the maximum 
magnitude of the latter is brighter by 0.5--1 mag.

 These objects also exhibit light variations called (positive) superhumps 
during superoutburst. The observed period of the superhumps 
is a few percent longer than the orbital period of the system.
The positive superhumps are thought to arise by periodic viscous
dissipation of tidally elongated disk (i.e. the eccentric disk)
whose aspsidal line slowly precesses in the prograde direction
(see \cite{whi88tidal}, \cite{hir90SHexcess}
)
 On the 
other hand, some cataclysmic variables show variations 
shorter than the orbital period called ``negative superhumps'' 
(\cite{uda88ttari};
\cite{har95v503cyg}; \cite{rin12v378peg}). 
 The origin of negative superhumps is usually considered as a 
result
of retrograde precession in a tilted accretion disk 
\citep{woo07negSH}.  When the disk is tilted,
the hot spot is formed at the inner part of the disk, not the
edge of the disk. Since the energy of hot spot comes from the
release of gravitational energy, such change in the location of
the hot spot causes a variation in the amount of the released energy, 
 namely the luminosity of the hot spot. 
Combined with the retrograde precession, this
effect can explain the negative superhumps.

The interval of successive two superoutbursts is called ``supercycle''.
These two superoutbursts usually sandwich several normal outbursts.
In order to explain such behavior of SU UMa-type dwarf novae, the tidal thermal
instability (TTI) model is suggested \citep{osa89suuma}. In this
model, systems with small mass ratios ($M_{2}/M_{1} = q \le 0.25$) enable
the disk to reach the radius of the 3:1 resonance to the orbital motion 
of the secondary. In normal outbursts, the material of the disk
only partly accrete to the inner region. The radius of the disk
becomes gradually larger after experiencing normal outburst. 
 When the disk radius
reaches the 3:1 resonance radius, the eccentric instability is excited.
The increased turbulence in the disk
causes an increase in the mass-transfer rate in the disk
and a long, bright superoutburst is triggered \citep{osa89suuma}.
This prograde precession also causes the superhump
(\cite{whi88tidal}, \cite{hir90SHexcess}). 

 ER UMa is a member of SU UMa-type dwarf novae and its 
intervals of
superoutburst (supercycle) are as short as
 40 -- 50 d \citep{kat95eruma}.
This object is known as the prototype of a subgroup,
``ER UMa type'' which is characterized by having extremely
short supercycles ($<$ 60 d)
 among SU UMa-type stars (e.g. \cite{rob95eruma}, 
\cite{nog95v1159ori}; for a
review see \cite{kat99erumareview}). 
Although \citet{gao99erumaSH} and \citet{kju10eruma} suggested on 
the presence of negative
superhumps during quiescence and a normal outburst, only positive 
superhumps were observed during the following superoutburst.
 However, \citet{ohs12eruma} reported that negative superhumps 
were detected in ER UMa during the superoutburst in 2011 January. 
This is the first
confident detection of negative superhumps during the superoutburst
of any SU UMa-type dwarf nova. In \citet{ohs12eruma} (hereafter paper I),
we reported the persistent negative superhump was detected in
ER UMa and implied the possibility that the existence of negative
superhumps suppresses of the occurrence of normal outbursts
.
 
  The existence of negative superhumps during superoutburst was 
also reported in other SU UMa-type dwarf novae V1504 Cyg and V344 Lyr
(\cite{osa13v1504cygKepler}, \cite{sti10v344lyr}). \citet{osa13v1504cygKepler} analyzed 
data of an SU UMa-type dwarf nova V1504 Cyg observed by Kepler, 
and showed that this object also shows negative superhumps during 
superoutburst as well as during normal outbursts and in quiescence. 
\citet{osa13v1504cygKepler} have
demonstrated that in V1504 Cyg the frequency of occurrence of normal 
outbursts in a
supercycle is related to the existence or non-existence of 
negative superhumps in the
sense that it is reduced when the negative superhumps exist, 
confirming the suggestion
made in Paper I. That is, the cycle lengths of normal outbursts 
are longer in a
supercycle with negative superhumps than those in a supercycle
 without negative
superhumps, and they called the former case ``type L-supercycle'' 
and the latter case
``type S-supercycle'' (these two types were first introduced by 
\citet{sma85vwhyi} for
supercycles observed in VW Hyi and symbols ``L'' and ``S'' come 
from ``long'' and ``short'' for normal outburst cycles).

 In Paper I, we dealt with  only one supercycle  of ER UMa in 2011.
We have made further comprehensive observations of ER UMa in 2011 and 2012
and, our dat covered three supercycles in 2011 and three supercycles in 2012.
All of  supercycles observed have turned out to be were all accompanied with negative superhumps, and
 provide an excellent opportunity to study the  outburst
behavior when negative
and positive superhumps co-exist.
In this paper, we report on these new observations together with a
more sophisticated analysis of the data reported in paper I.
 We explain our  observations in section \ref{obschap} and we
present the result of their analysis
observation and analysis in section \ref{reschap}. The conclusion is
given in section \ref{conchap}.

\section{Observations}
\label{obschap}

\begin{figure}
   \begin{center}\label{chart}
     \FigureFile(70mm,70mm){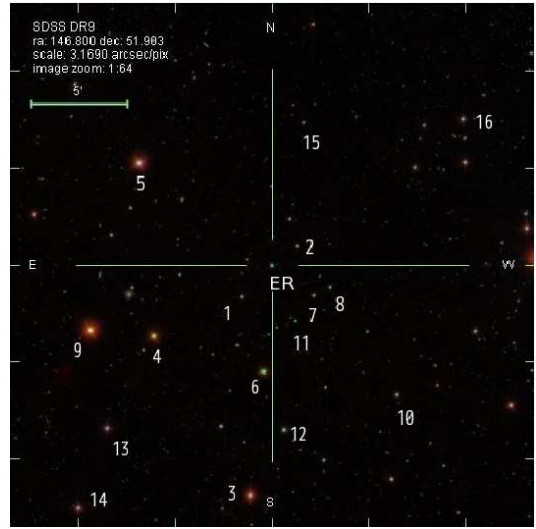}
   \end{center}
   \caption{Finding chart of ER UMa and the comparison stars. (derived from
http://cas.sdss.org/dr7/en/tools/chart/chart.asp)}
 \end{figure}

 We performed time-resolved photometric observations of ER UMa
 at 21 
observatories scattered world-wide from 2011 
January to 2012 December as a part of the VSNET Collaboration 
\citep{VSNET}. We could obtain data of 146 nights in 2011, and 
161 nights in 2012. The log of observatories is given in
table \ref{obslist}, where (1) the first column is the
abbreviation of observer, (2) names of observer or observatory,
(3) instrument used, and (4) comparison starts used. The journal
of observations is summarized in table \ref{tab:log} (given Appendix 
in the Electric version). The chart of ER UMa and its comparison 
stars is presented in figure 1.

 After dark-subtracting and flat-fielding in CCD observations, we 
performed 
aperture photometry of the variable and its comparison stars 
and obtained differential magnitudes. 
All observed times were transformed to barycentric Julian 
Days (BJD). We made corrections for the systematic differences 
between observers after that.

\begin{longtable}{cccc}
\caption{Log of observatories}\label{obslist}

\hline\hline
The key to & Observer or  & Instrument & Comp \\

Observer &Observatory Name & & \\
\hline
\endhead
\hline
\multicolumn{4}{l}{1: GSC 3439.629, 2: GSC3439.920 3: TYC2-3439.1287.1 4: TYC2-3439.1099.1 5: TYC3439.1253.1 } \\
\multicolumn{4}{l}{6: GSC3439.669 7: GSC3439.816 8: GSC3439.957 9: TYC3439.1211.1 10: GSC3439.745 }\\
\multicolumn{4}{l}{11: USNO1350.07816004 12: GSC3439.1105 13: GSC3439.911 14: TYC2-3439.916.1}\\
\multicolumn{4}{l}{15:GSC3439.1091 16: TYC2-3439.885.1}\\
\multicolumn{4}{l}{\commenta by F. J. Hambsch and T. Krajci}\\
\multicolumn{4}{l}{\commentb by Natalia Katysheva} \\
\hline\hline
\endfoot
\hline
\multicolumn{4}{l}{1: GSC 3439.629, 2: GSC3439.920 3: GSC3439.1287 4: TYC2-3439.1099.1 5: TYC3439.1253.1 } \\
\multicolumn{4}{l}{6: GSC3439.669 7: GSC3439.816 8: GSC3439.957}\\
\multicolumn{4}{l}{9: GSC3439.1211 10: GSC3439.745 11: USNO1350.07816004}\\
\multicolumn{4}{l}{12: GSC3439.1105 13: GSC3439.911 14: TYC2-3439.916.1}\\
\multicolumn{4}{l}{15:GSC3439.1091 16: GSC3439.885}\\
\multicolumn{4}{l}{\commenta by F. J. Hambsch and T. Krajci}\\
\multicolumn{4}{l}{\commentb by Natalia Katysheva} \\
\hline\hline
\endlastfoot
KU & Kyoto University & 40cmSCT$+$ST$-$9E & 1 \\
Aka  & Akazawa Hidehiko&  28cmSC,35.5cmSC  & 1,4,5 \\
 & & $+$ ST-7XE, ST-9XE & \\
AKz & Astrokolkhoz team\commenta  & 30cmSC$+$ST$-$9 ,35cmSC$+$ST$-$8 & 2 \\
APO & Apache Point Observatory & 50cmC+SITe & 6 \\
BBo & Boyd Boitnott &28cmSC$+$QSI$-$516wsg & 6\\
CRI & Crimean Astrophy. Obs. &60cm$+$Apogee Alta E47 & 6  \\
deM & Enrique de Miguel & 28cmSC$+$QSI$-$516wsg / 25cmL & 2\\
DPV & Pavol A. Dubovsky & 28cmL$+$DSI ProII & 6  \\
Ham & Frantz$-$Josch Hambsch & 40cm+STL11 & 2 \\
Ioh & Itoh Hiroshi & 30cmSC+DSI-Pro­¶ & 6 \\
IMi & Ian Miller & 35cmSC$+$SXVR-H16 & 1,2,11\\
Kai & Kasai Kiyoshi  & 28cmSC+ST-7XME & 1,4\\
Kra & Tom Krajci &28cmSC+SBIG ST-8& 6  \\
LCO & Colin Littlefield & 28cmSC+ST$-$8XME & 5,13,14\\
Mhh & Hiroyuki Maehara  &25cmL+ST$-$7XME & 4  \\
NDJ & Nick James &28cmSC$+$SBIG ST-9XE & 2  \\
NKa & Natalia Katysheva & 50cmR, 14cmC+ST-10XME&  1,2,4,6, \\
 & & & 8,9,10,12  \\
OKU & Osaka Kyoiku Univ. & 51cm$+$ST$-$10 & 1 \\
OUS & Okayama Univ. of Sci. team & 23.5cmSC+ST$-$8 & 6 \\
PSD & Stefano Padovan &25cm epsilon+ST-10XME & 1,3,8, \\
& & & 10,13,15  \\
Pol & Polaris Observatory & ST$-$7E & 4 \\
Rui & Jevier Ruiz & 0.4mRC+ST$-$8XME & 1\\
Sac & Seikei High School  & 15.2cmR$+$ST$-$9E& 5\\
Shu & Sergey Shugarov &50cmR, 14cmC+ST-10XME & 1,2,4,6, \\
  & & & 8,9,10,12  \\
Siz & Siokawa Kazuhiko & 35SC+ST-9E & 4 \\
SAO & Special Astrophysical & 1m+EEV CCD42-40' & 1,2,4,6, \\
 & Observatory\commentb  & & 8,9,10,12    \\
SWI & William Stein & C14$+$SBIG ST-10XME & 2 \\
Ter & Terskol Observatory &C14$+$STL1001  & 6  \\
Vir & Jani Virtanen &C14$+$SBIG ST-10XME & 6  \\
VIR & Natalia Virnina & 60cm & 6  \\
Vol & Irina Voloshina & 60cmL$+$Apogee 47&  11 \\
\end{longtable}

\section{Result}
\label{reschap}

\begin{table}
\begin{center}
\caption{List of superoutbursts}\label{obslog1}
\begin{tabular}{ccccc}
\hline\hline
ID & The starting date of & The maximum date of  & The maximum & The length of \\ 
 &  superpoutburst (BJD-2400000) & (BJD-2400000 )  & magnitude & supercycle (d) \\
\hline
2011 S1 &   55578\commenta & - & 12.6\commenta & $-$ \\
2011 S2 &   55622.0 & 55625.1 & 12.7  &  44  \\
2011 S3 &    55671.9 & 55674.4 & 12.7 &  50 \\
2012 S1 &  55927\commentd & 55929\commentd  & 12.9  &    58\commentb \\
2012 S2  &    55981.3 & 55982.6 & 13.0 &  54     \\
2012 S3 &    56033.8  & 56034.4  &12.9 &   53   \\
2012 S4  & - &  56088.9\commentc& - & 55   \\
\hline
\multicolumn{5}{l}{\commenta Based on VSNET data}\\
\multicolumn{5}{l}{\commentd The precise timing is unclear because of the scarcity of observations)}\\
\multicolumn{5}{l}{\commentb The date of previous superoutburst is based on VSNET data.}\\
\multicolumn{5}{l}{\commentc The estimated time of maximum, }\\
\multicolumn{5}{l}{not the start of the outburst (due to the scarcity of observations)} \\
\hline\hline
\end{tabular}
\end{center}
\end{table}

\begin{longtable}{cccc}
\caption{List of normal outbursts}\label{nobslist}

\hline
\hline
ID & Cycle length\commente &The starting date of outburst & Maximum\\
& (d) & (BJD$-$2400000) & Magnitude\\
\hline
\endhead
\hline
\multicolumn{4}{l}{\commente The cycle length from the previous outburst to the}\\
\multicolumn{4}{l}{current one} \\
\multicolumn{4}{l}{\commenta The estimated maximum time, not the start of} \\ 
\multicolumn{4}{l}{the outburst (due to the scarcity of observations)} \\
\multicolumn{4}{l}{\commentb Precursor outburst of the next superoutburst} \\
\multicolumn{4}{l}{\commentc Suspected superoutbrust} \\
\multicolumn{4}{l}{\commentd Not so confirmed} \\
\hline
\hline
\endfoot
\hline
\multicolumn{4}{l}{\commente The cycle length from the previous outburst to the}\\
\multicolumn{4}{l}{current one} \\
\multicolumn{4}{l}{\commenta The estimated maximum time, not the start of} \\ 
\multicolumn{4}{l}{the outburst (due to the scarcity of observations)} \\
\multicolumn{4}{l}{\commentb Precursor outburst of the next superoutburst} \\
\multicolumn{4}{l}{\commentc Suspected superoutbrust} \\
\multicolumn{4}{l}{\commentd Not so confirmed} \\
\hline
\hline
\endlastfoot
2011 N1-1 & - &  55592.2 & 13.6 \\
2011 N1-2 & 5.7  &  55597.9  & 13.5 \\
2011 N1-3 & 6.6 &  55604.5  & 13.4 \\
2011 N1-4  & 6.6 &  55611.1  & 14.3  \\
2011 N1-5 &  4.0  &  55615.1 & 13.3  \\
2011 N2-1 &  - &  55644.7  & 13.6 \\
2011 N2-2 &  5.5 &  55650.2 & 13.4  \\
2011 N2-3 &  6.7  & 55656.9  & 13.5 \\
2011 N2-4 &  7.2 &  55664.1 & 13.2  \\
2011 N3-1 &  -  &  55693.3 & 13.7   \\
2011 N3-2 & 4.8 &   55698.1  & 13.6  \\
2011 N3-3 &  7.4 &   55705.5 & 13.6\commentd  \\
2011 N3-4   &  6.8 &   55712.3  & 13.4   \\
2011 N3-5\commentc &  9.0\commentd &  55721.3  & 13.2\commentd \\
\hline
2012 N0-1 &  - &   55900.5  & 13.6  \\
2012 N0-2 & 9.0 &   55909.5  & 13.5 \\
2012 N0-3  &9.5  &  55919.0  &  13.4\\
2012 N1-1 &  - &   55946.3   &  13.9 \\
2012 N1-2 & 5.9  &  55952.2 & 13.7  \\
2012 N1-3 & 8.9   &  55961.1   & 13.8  \\
2012 N1-4 &  8.0 &   55969.1  &  13.5 \\
2012 N2-1 &  - &   55998.8  & 13.5  \\
2012 N2-2 &  5.9 &   56004.7  & 13.9  \\
2012 N2-3 &  8.2 &   56012.9  &  13.9 \\
2012 N2-4 &  9.3 &   56022.2  & 13.5 \\
2012 N2-5 &  9.4  &  56031.6\commentb   & 13.5  \\
2012 N3-1 &  -  & 56051.0\commenta & 13.7\\
2012 N3-2 &  6.4 &   56057.4\commenta   & 13.8  \\
2012 N3-3 &  6.5 &   56063.9\commenta  & 13.7   \\
2012 N3-4 &  6.7  &   56070.6\commenta  & 13.7 \\
2012 N3-5 &  11.3 &   56081.9\commenta  & 14.2\\ 
\end{longtable}

\begin{figure}
   \begin{center}
     \FigureFile(120mm,180mm){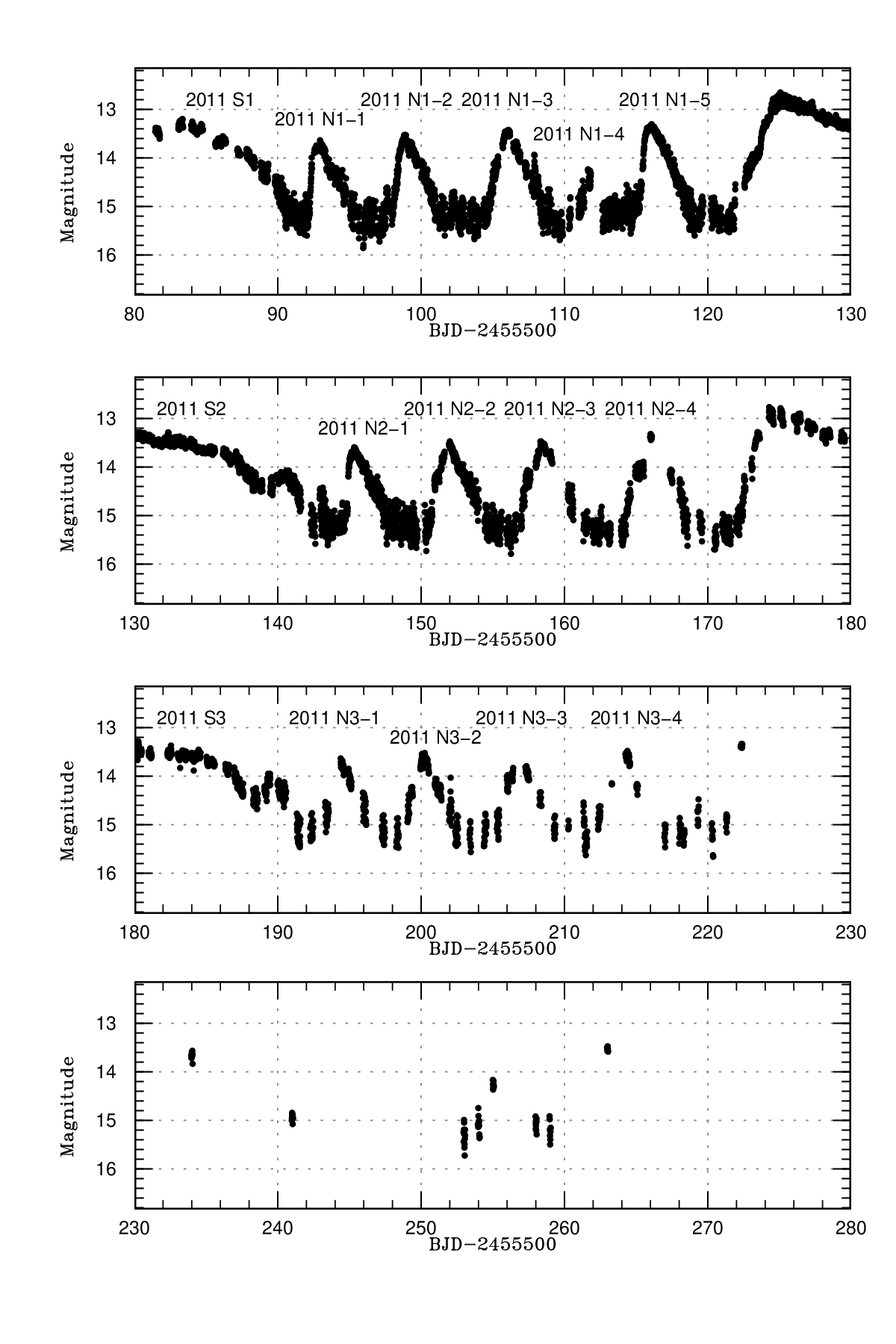}
   \end{center}
   \caption{Entire light curve of ER UMa of the 2011 season. The observed data is binned to 0.01 d.}\label{wholelc}
 \end{figure}

\begin{figure}
   \begin{center}
     \FigureFile(120mm,180mm){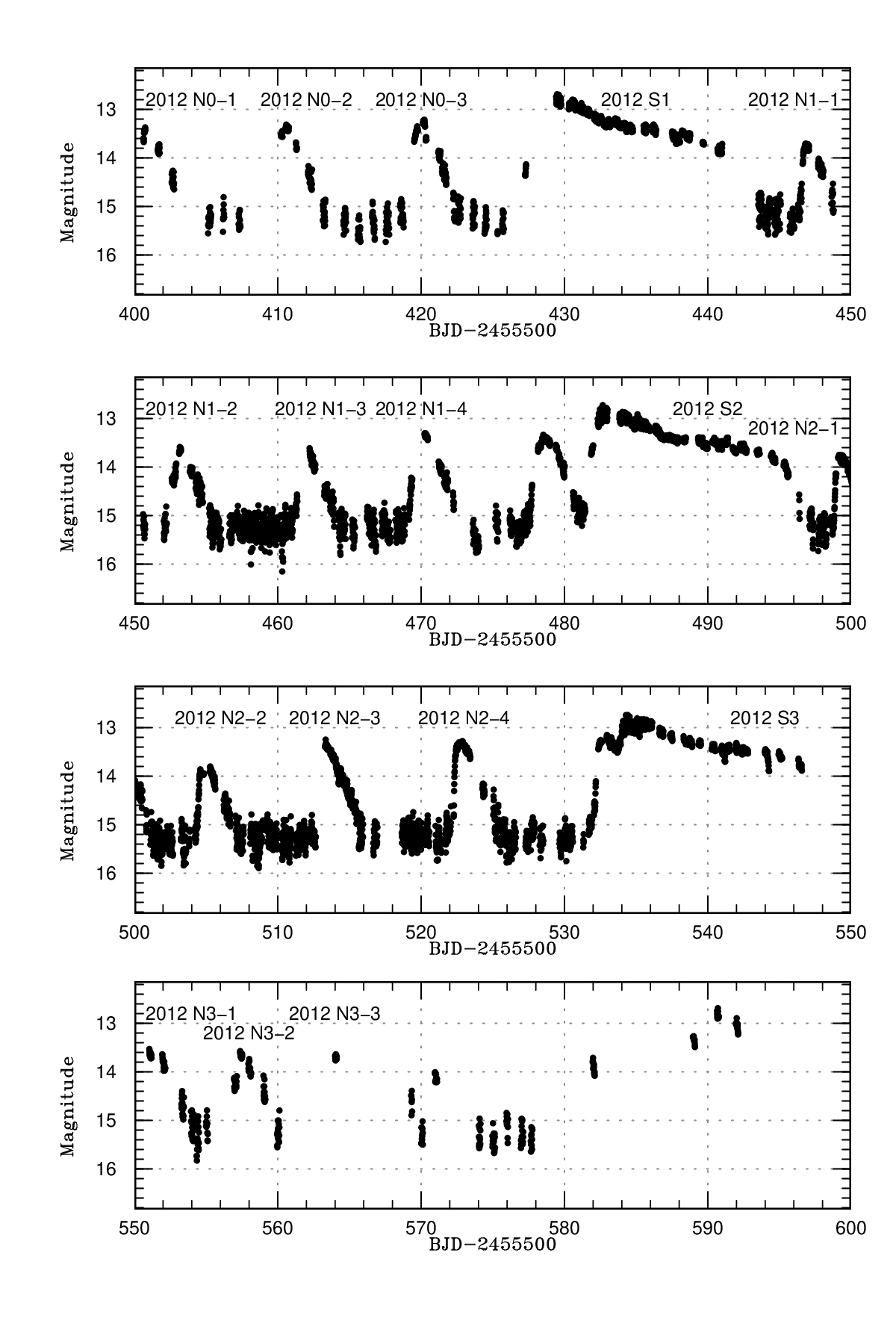}
   \end{center}
   \caption{Entire light curve of ER UMa of the 2012 season. The observed data is binned to 0.01 d.}\label{wholelc2}
 \end{figure}

In this section, we first present the results of our observations
and their analysis then discuss their implications. 
We describe first the outburst behaviors of ER UMa
 in subsection 3.1, and we then discuss the negative superhumps
and the positive superhumps in subsection 3.2, and 3.3, respectively,
and we finally deal with the transition from negative superhump
to positive superhump in subsection 3.4.

 \subsection{Outburst Light Curves}

\subsubsection{The overall light curves of ER UMa in 2011 and 2012}
 The overall light curve of ER UMa is presented in figures 
\ref{wholelc} for 2011 and \ref{wholelc2} for 2012. 
 The entire list of observed superoutbursts is shown in 
table \ref{obslog1},  where (1) the first column is the identification
of the superoutburst, (2) its starting  date in BJD, (3) The date of BJD of the supermaximum,
and (4) the length of supercycle, where the supercycle is defined from the
starting date of the previous superoutburst to that of the current
one. As seen in the figure \ref{wholelc} and \ref{wholelc2},
 three superoutbursts in 2011 (2011 S1-S3) and three superoutbursts 
in 2012 (2012 S1-S3) were clearly observed during this observational 
campaign although a possible superoutburst candidate was also 
observed in 2012 (2012 S4) but it was unclear because of sparse
data points. The length of observed  of supercycles were 
in a range of 44--58 d. These values are in agreement with the 
previous report \citep{zem13eruma}.
 Table \ref{nobslist} shows the list of normal outbursts. The first
column is the identification of a normal outburst. Here 2012
N2-3, for instance, indicates the third normal outburst in supercycle
No.2 in 2012. We see from table \ref{nobslist} that the maximum magnitude
of normal outburst gets brighter and the length of normal outburst
cycles gets longer as the object approaches the next superoutburst
(i.e. with an advance of supercycle phase), although there
exists some exceptions (e.g. the interval between 2012 N1-2 
and 2012 N1-3 is longer than that between 2012 N1-3 and 2012 N1-4).

\subsubsection{Description of individual supercycles}

 Let us now examine the outburst behavior of individual supercycles.
Here we define the identification of a supercycle by that of the
starting superoutburst in a supercycle, that is, ``supercycle 2011 S1'',
for instance, is a supercycle which begins with a superoutburst
2011 S1,  and ends with 2011 S2. The duration of a superoutburst
is defined as a period from the time of maximum of the superoutburst
to its end because the observations of rising stage were lacking
in some cases.

 Supercycle 2011 S1:

 The superoutburst 2011 S1 was the first superoutburst when 
negative superhumps were first detected (Paper I). 
Time resolved observations started three days after the detection of the 
superoutburst. 
 Unfortunately the rising part of this superoutburst was not observed 
and the existence of the positive superhumps in this
superoutburst was not confirmed. However, we suspect that the 
positive superhumps must have appeared most likely in the earliest
 phase of this superoutburst.

The superoutburst lasted for probably 13--15 d. 
Two days after the end of the superoutburst, the next normal 
outburst (2011 N1-1) started.
 In this supercycle, this object showed four normal outbursts
 (BJD 2455592, 2455597, 2455604, and 2455615). Besides
them,  a mini-outburst with the amplitude of only 1 mag occurred 
(BJD 2455611). The length of the supercycle was 44 d.

 Supercycle 2011 S2 and 2011 S3:

 The superoutburst 2012 S2 started as a form of a normal 
outburst and the start of the outburst was BJD 2455622
and persisted for 16 d. On BJD 
2455642, 14 d after the supermaximum, the declining of the brightness
temporarily ceased and this object
brightened by 0.5 mag. The superoutburst 2011 S3 is similar
to 2012 S2.

 Supercycle 2012 S0:

 Three normal outbursts (BJD 2455900, 2455909, 2455919) 
were caught before the first superoutburst where the 
time resolved observations were performed in the 2012 
season. According to monitoring observation reported 
to VSNET, the previous superoutburst (2012 S0) occurred 
around BJD 2455869. 

 Supercycle 2012 S1:

 Because of the lack of observations, it is unclear 
when the superoutburst 2012 S1 decayed although the 
decline occurred between BJD 2455941 and 2455943. 
Considering the maximum of the superoutburst was 
around BJD 2455962, the duration of superoutburst was 
approximately 15 d.

\begin{figure}
   \begin{center}
     \FigureFile(160mm,80mm){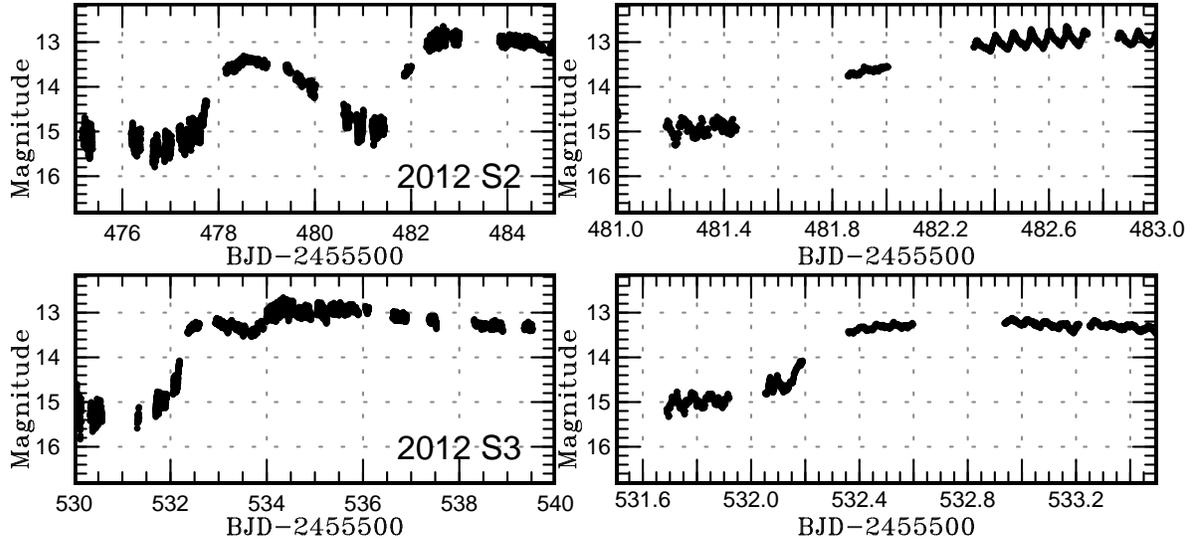}
   \end{center}
   \caption{Enlarged light curve of the rising stage
 of 2012 S2 (upper two panels) and 2012 S3 (lower two 
panels). The observation point is binned by 0.001 d.
The left two panels show the precursors and the start 
of superoutbursts. And right two panels is further 
enlarged light curve in order to show the superhump 
variations. The positive superhumps became dimmed and 
the positive superhump evolved. }
\label{zoomedlc}
 \end{figure}

 Supercycle 2012 S2:

The superoutburst 2012 S2 is an interesting case since 
the superoutburst was triggered on the way of decaying 
of normal outburst (the upper diagram of figure \ref{zoomedlc}).
 The duration of the superoutburst 2012 S2 was 14 d. 
 After the superoutburst ended, 
this object showed four normal 
outbursts (BJD 2456051, 2456057, 2456063, 2456070, 
2456081) occurred.

 Supercycle 2012 S3:

 The superoutburst 2012 S3 is a typical superoutburst, 
which has a ``shoulder'' at the beginning of the 
superoutburst (the lower diagram of figure \ref{zoomedlc}), 
which was shown in \citet{osa13v1504cygKepler}.
 Despite of the lack of the time-resolved observations 
in the late stage of the superoutburst 2012 S3, the 
VSNET data shows that this object declined around BJD 
2456048--2456049. Thus  the duration of the superoutburst 
2012 S3 was 14--15 d.
After the superoutburst decayed, four or five normal 
outbursts (BJD 2456051, 2456057, 2456063, 2456070, 
and 2456081) are detected. However, the fifth normal
 outburst (BJD 2456081) may be a precursor outburst of 
the next superoutburst 2012 S4. However, at any rate, 
the definite property of these outbursts are not clear 
because of the lack of the observations. The property 
of next outburst (2012 S4) is also unclear for the same 
reason.

\subsection{The Frequency of Normal Outbursts}

 The number of normal outbursts
in a supercycle  in 2011 and 2012 was found to be mostly
four and sometimes five (table \ref{obslist}). 
The cases of five are exceptional and they will be discussed
below.  We reported in Paper I that the frequency of normal outbursts
in the first supercycle of 2011 (i.e., 2011 S1 ) as lower than those found
in previous observations of ER UMa and we suggested that the 
existence of negative
superhumps (and so the existence of a tilted disk) might 
suppress frequent occurrence
of normal outbursts. \citet{zem13eruma} pointed out that 
the number of normal
outbursts in a supercycle of ER UMa varied between 4 and 6 
for a long time scale and
they also suggested that these differences are related to 
the appearance of the
negative superhumps. As discussed in the next subsection,

 it has turned out that all
supercycles of ER UMa observed in 2011 and 2012 were accompanied by
the negative superhump and thus they were the type L-supercycles
(see, \citep{osa13v1504cygKepler}, for the definition of the type L- 
and type S-supercycles).

This trend did not change largely between 
two seasons. Although five normal outbursts were observed  
between 2012 S1 and S2, the superoutburst 2012 S2 occurred at 
the declining stage of the fifth normal outburst. Thus this 
normal outburst was a precursor of the next superoutburst. 
Another exception is 2011 N1-4, which will be discussed in 
the later subsection. Except for these cases, four normal 
outbursts were observed during one supercycle.
The same correlation
between the appearance of negative superhumps and the frequency
 of normal outburst
in a supercycle was known in other SU UMa stars, 
V503 Cyg (\cite{kat02v503cyg}, \cite{pav12v503cyg}) and 
V1504 Cyg \citep{osa13v1504cygKepler}. In this respect, 
Two exceptional cases of five normal outbursts in a supercycle are discussed
here. As already mentioned in the individual supercycles, five normal
outbursts occurred in supercycle 2012 S1 but the fifth one has turned 
out to be a precursor normal outburst of superoutburst 2012 S2 and 
it may be regarded as a part of the next superoutburst. Another 
exception is 2011 N1-4 and it was found to be a
mini-outburst and it will be discussed in the later subsection.

\subsubsection{Light Curve Profile and Rising Rate of 
Normal Outbursts in a Supercycle}

\begin{figure}
  \begin{center}
    \FigureFile(140mm,210mm){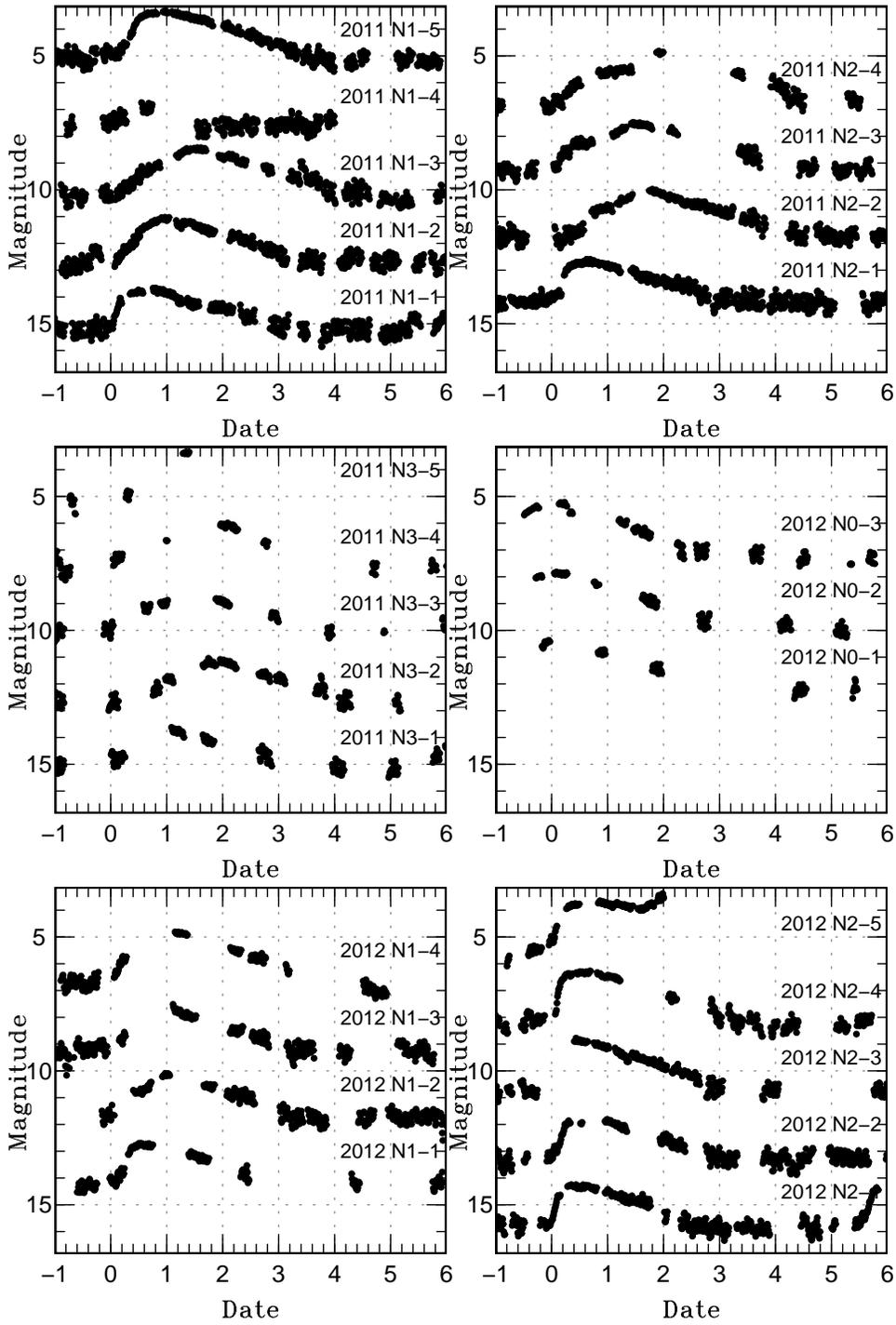}
  \end{center}
  \caption{Rising stage of normal outbursts. In the cycle 
of 2012 N0, the maximum of normal outbursts is set at zero, 
and in other cycles the start of normal outbursts is set 
at zero.}
  \label{fig:rise}
\end{figure}

\begin{figure}
   \begin{center}
     \FigureFile(140mm,210mm){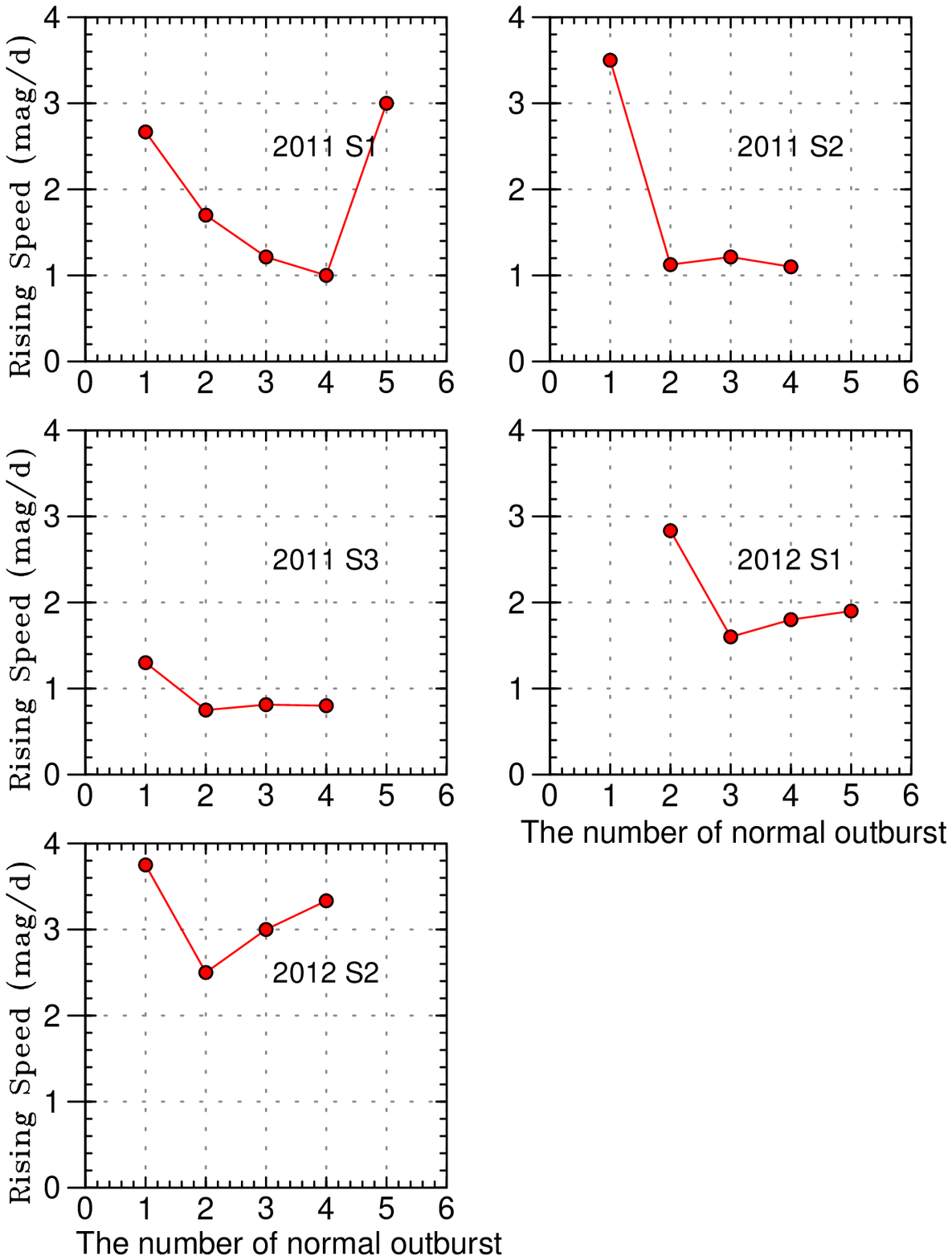}
   \end{center}
   \caption{Rising speed variation of each supercycle. 
Since the rising speed can be estimated because of the 
unclear timing of the start of outburst in some of normal 
outbursts, they are omitted. }\label{rises}
 \end{figure}

 Figure \ref{fig:rise} illustrates variation in 
light curve profile of normal outbursts within
supercycles. Figure \label{rises} exhibits the 
rising rates of normal outbursts within each
supercycle. In the case of supercycle 2011 S1 (the 
left top panel of figure \ref{fig:rise}), we
see that the rising rate to the maximum was getting 
slower with advance of supercycle
phase except for the fifth one. We find from figure \ref{fig:rise}
that, generally speaking, the rising rate of the first normal 
outburst within each supercycle was faster than those
is the light curves during of later ones.
Two types of profiles in outburst light-curves are known 
to exist. The first one is that its rising rate is faster 
than its declining rate (i.e., rapid rise and slow
decline) and the other one is that the rising rate to the 
maximum is not so fast but
more or less similar to the decline rate (i.e., the light 
curve profile is more or
less symmetric with respect to the rise and fall). In 
the disk instability model for
the outburst of dwarf novae, the former type of light 
curve is produced by the
``outside-in'' type outburst (or the type A outburst in
\citet{sma84DI}) in which the transition to the hot 
state starts from the outer-part of the accretion
disk and the heating front propagates inward while the 
latter type of outburst is produced by the ``inside-out'' 
outburst (type B in \citet{sma84DI}) in which the transition to
the hot state starts from the inner-part of the accretion disk.
We find from these two figures that the first normal 
outburst in a supercycle looks
like an ``outside-in''-type, while most of other outbursts 
do like ``inside-out''-type.

\subsection{Negative superhumps}


 As shown in Paper I, negative superhumps are 
detected in ER UMa during superoutburst as well as during 
normal outbursts and quiescence. The negative superhumps were clearly 
detected except for the early stage of the superoutburst. 
Their amplitude is 0.5--1.0 mag in quiescence. 
 We also show 
the amplitude variation in flux unit in figures \ref{oc11}, 
\ref{oc12}. In these diagram, 1 magnitude variation in 16 
mag is normalized as 1. No dramatic change was associated 
with the change between quiescence and outburst is seen 
in flux unit (figures \ref{oc11}, \ref{oc12}) as already 
pointed in \citet{osa13v1504cygKepler}.


 The valid interpretation for negative superhumps is the tilted
accretion disk. The tilted disk show the retrograde precession,
and 
 \citet{lar98XBprecession} presented a equation of $q$ and 
$\epsilon_{-}$ for a retrograde precession of the tilted disk. 
According to \citet{lar98XBprecession},
the frequency of 
negative superhump frequency is given by

\begin{equation}
\epsilon^{*}_{-} = \frac{\nu_{\rm{orb}}-\nu_{\rm{nSH}}}{\nu_{\rm{orb}}}
 = -\frac{3}{7}\frac{q}{\sqrt{1+q}}(\frac{R_{\rm{d}}}{A})^{3/2} 
\rm{cos}\theta
\end{equation},
where $\nu_{\rm{orb}}$  and $\nu_{\rm{nSH}}$ are the frequency 
for negative superhumps and the orbital frequency of the 
binary, and $\theta$ is the tile angle of the disk to the 
binary orbital plane. If $\theta$ is small, we can assume  
$\rm{cos} \theta \sim 1$ and $\nu_{\rm{nSH}}$ can be determined 
by the disk radius $R_{\rm{d}}$ for specific system because 
$q$ and $A$ do not change in observational time-scale. 
 In a real disk,  $|\epsilon_{-}|$ represents the precession of
 the disk as a whole, to which precession rates from different
 radii contribute. Since the precession rates
 in smaller radii is smaller, the actual |$\epsilon_{-}$| is
 smaller than what is expected for a ring in the outer radius
 of the disk. For more details, see Appendix in
 \citet{osa13v344lyrv1504cyg}.

Indeed \citet{osa13v344lyrv1504cyg} indicated that the frequency of
negative superhump is useful probe to study the
change of the radius of accretion disks in SU UMa-type dwarf novae
through the analysis of V1504 Cyg. In this case,the frequency of negative
superhumps varies systematically during supercycles and the variation
of the frequency is a good probe of the variation of the disk. Now
we can observe the persistent negative superhump of ER UMa. Thus
We have a good opportunity to investigate the variation of the
disk radius through the variation of negative superhump period. 

We analyzed the period of negative superhumps in two methods.
In subsubsection 3.3.1, we will take the periodic analysis with a 
traditional method in the research of dwarf novae, drawing $O-C$ 
diagrams. And in subsubsection 3.3.2, we will adopt a new method called
least absolute shrinkage and selection operator (Lasso, \cite{lasso})
and perform two dimensional spectral analysis.

\subsubsection{$O-C$ analysis}

\begin{figure}
     \FigureFile(60mm,90mm){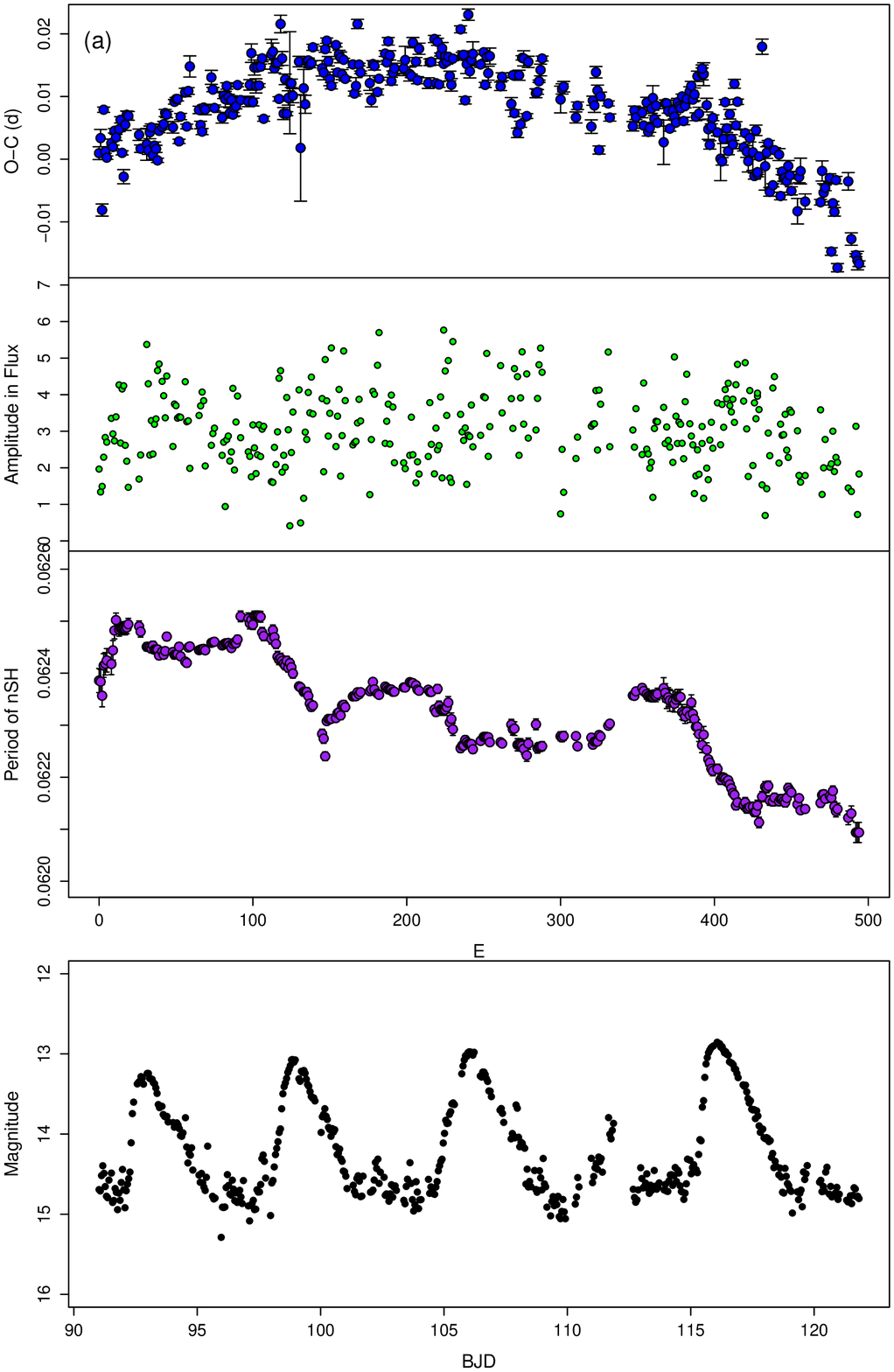}
    \FigureFile(60mm,90mm){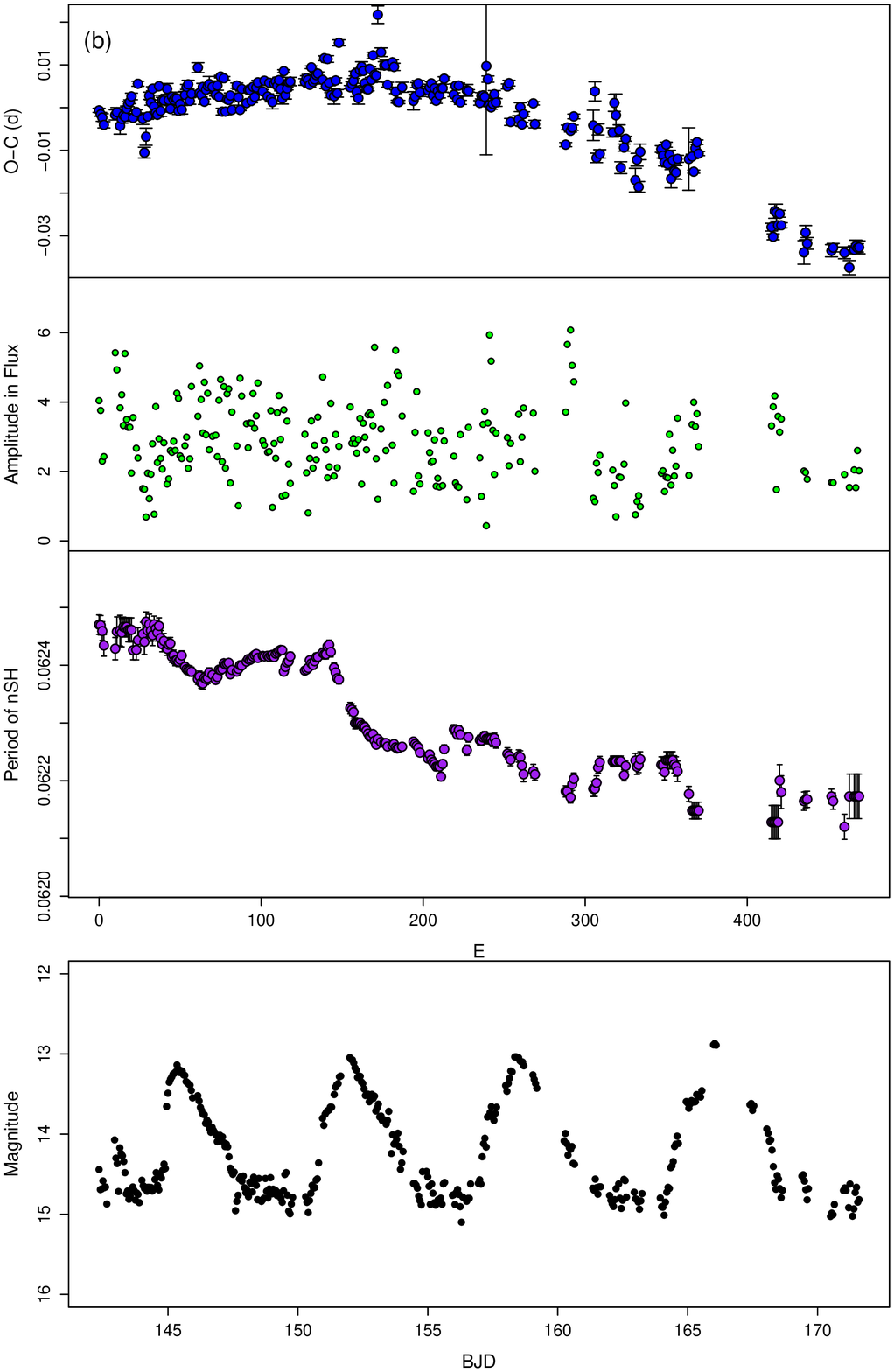}
\FigureFile(60mm,90mm){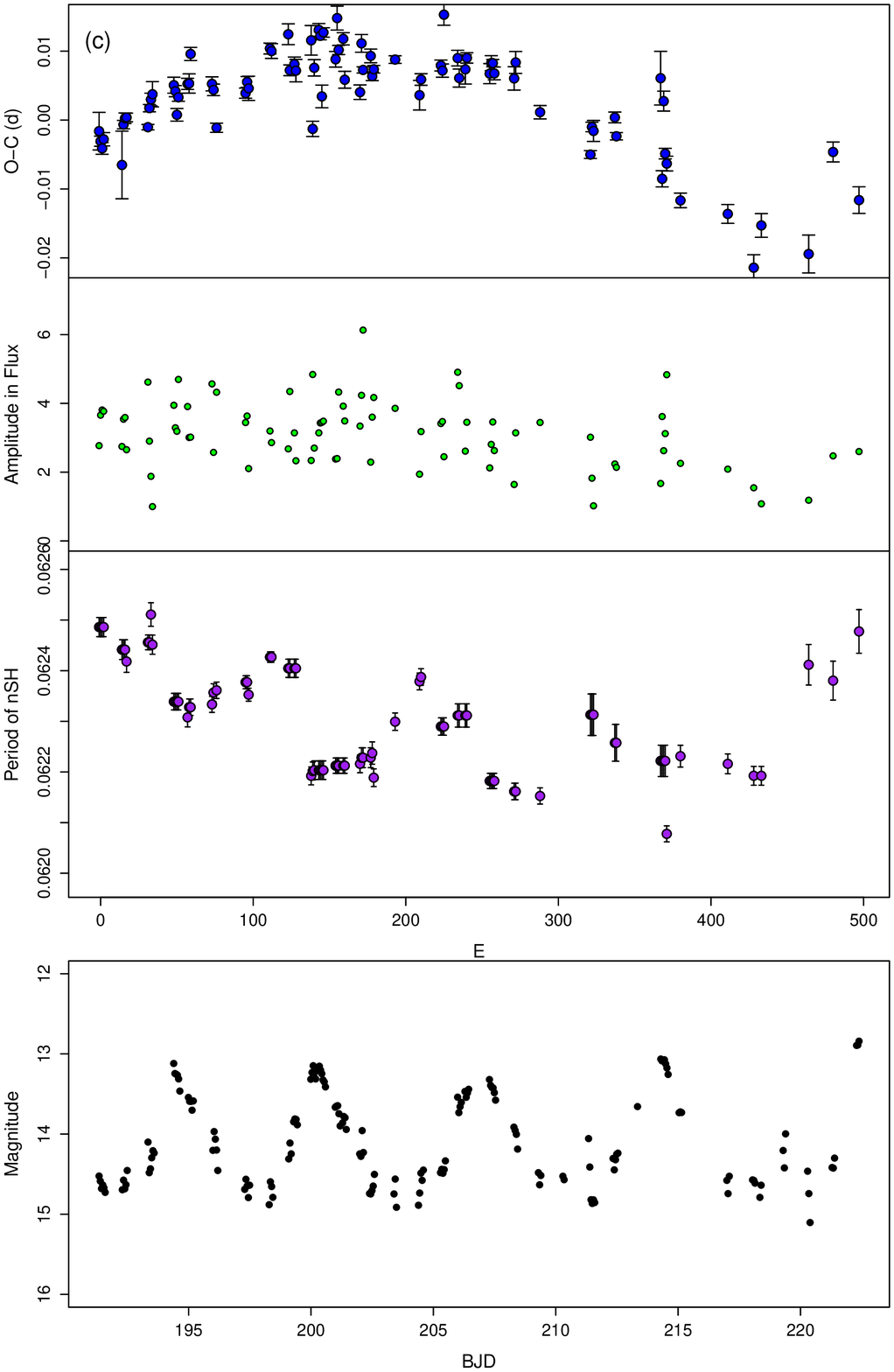}
  \caption{$O-C$ diagram of negative superhumps, and related diagrams of 
during 
each supercycle of 2011. The panel of (a), (b), (c) correspond to supercycle
2011 S1, 2011 S2, and 2011 S3. For each panel, top to bottom:
(1) $O-C$ diagram of negative superhumps, (2) The amplitude of negative
 superhumps in flux, (3) The period of negative superhumps estimate
 by PDM analysis, (4) The light 
curve. The $O-C$ value is against the 
equation of $2455591.020 + 0.062340 E$. The middle panel 
shows diagrams during 2011 S2-S3. The $O-C$ value is 
against the equation of $2455642.346 + 0.062339 E $. 
The lower panel shows diagrams between 2012 S3-4. The 
value is against the equation of $2455691.405 + 0.062305 E$}
\label{oc11}
 \end{figure}

\begin{figure}
    \FigureFile(60mm,90mm){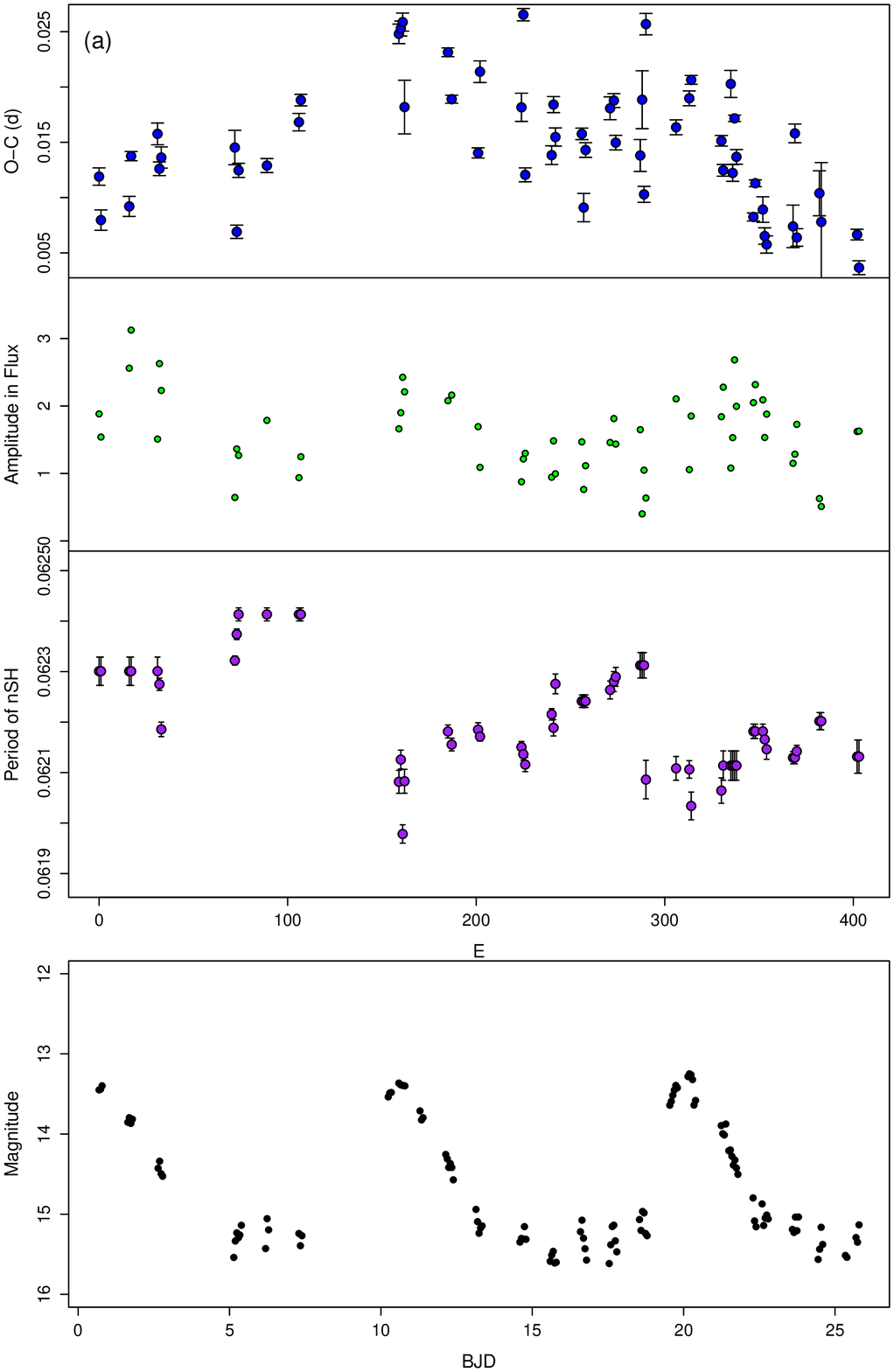}
    \FigureFile(60mm,90mm){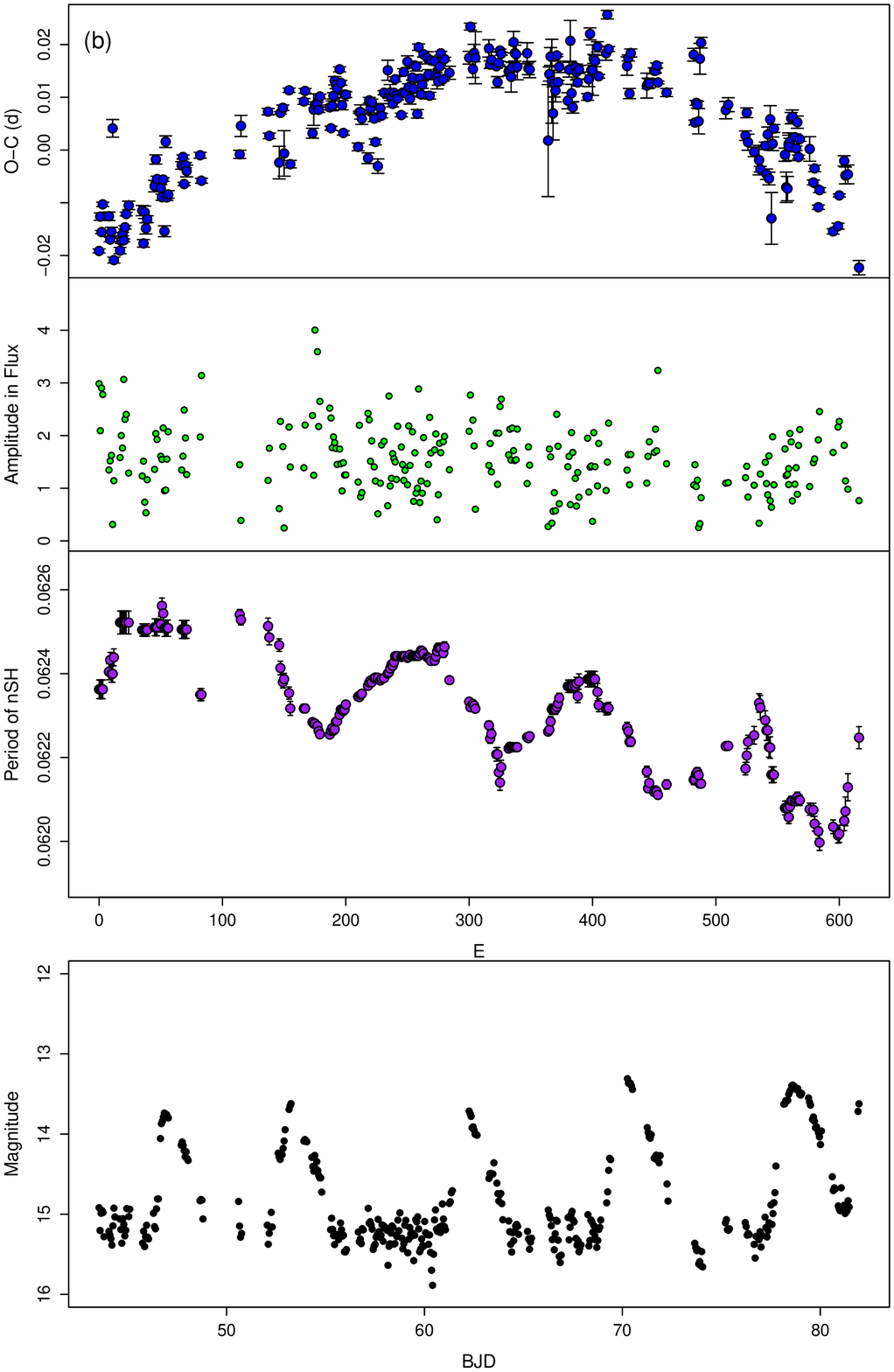}
\FigureFile(60mm,90mm){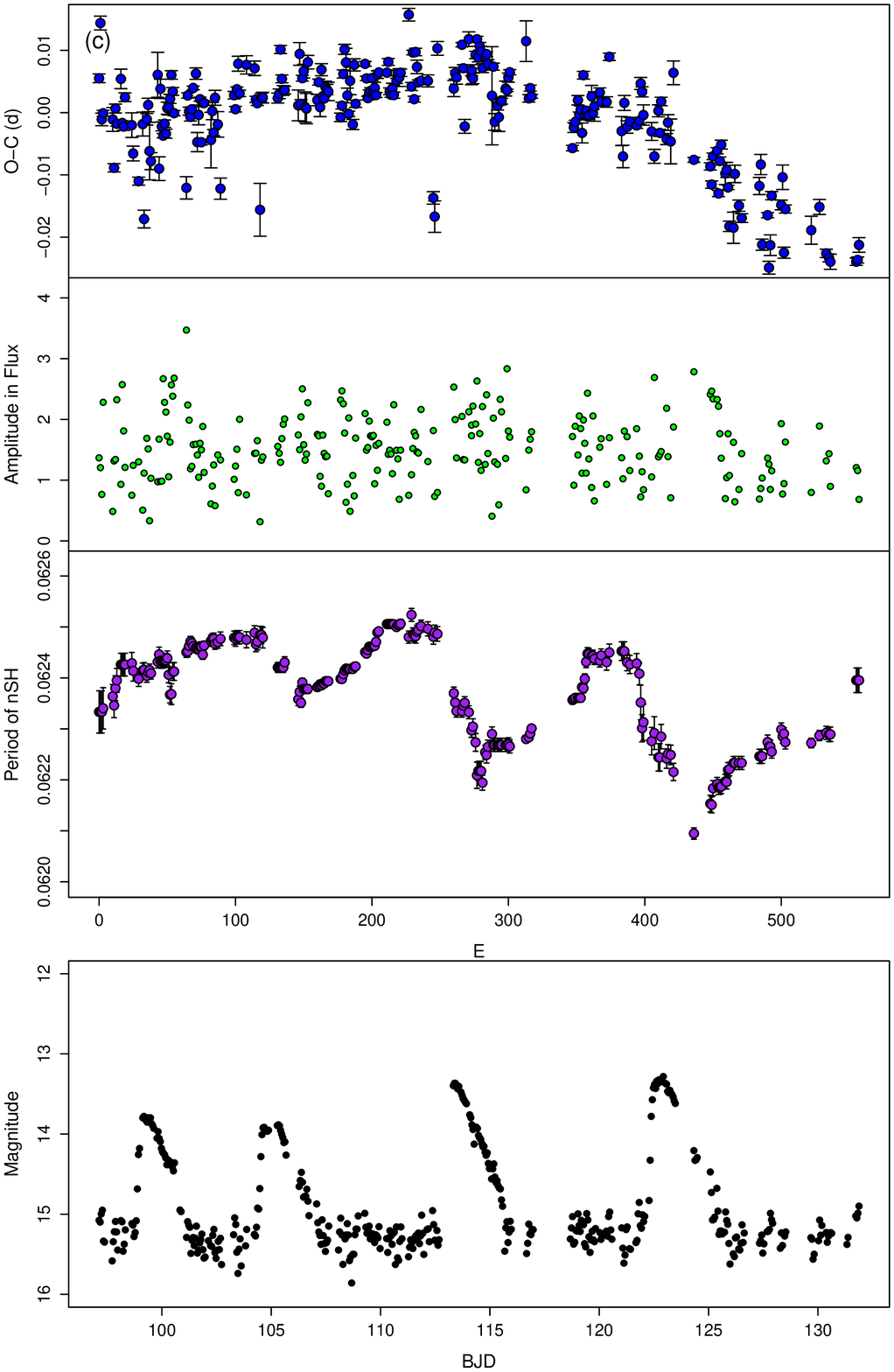}
  \caption{
$O-C$ diagram of negative superhumps, and related diagrams of 
during 
each supercycle of 2012. The panel of (a), (b), (c) correspond to supercycle
2012 S0, 2012 S1, and 2012 S2. For each panel, top to bottom:
(1) $O-C$ diagram of negative superhumps, (2) The amplitude of negative
 superhumps in flux, (3) The period of negative superhumps estimated by PDM analysis, (4) The light 
curve.
 The $O-C$ value is against the 
equation of  $2455900.680 + 0.06223  E$. The upper right 
panel shows diagrams during supercycle 2012 S0. The $O-C$ value 
is against the equation of $2455943.576 + 0.0623 E$. The 
lower panel shows diagrams between 2012 S1 and 2012 S2. 
The value is against the equation of $2455997.123 + 0.0624 E$.}
 \label{oc12}
\end{figure}

 We estimated the maximum timings of negative superhumps
 in the way given in \citet{Pdot} for data except for 
superoutburst stage.  The template for fitting the maximum 
was an average profile of negative superhump from data in 
quiescence between 2011 S1 and S2 (The period of used data 
for template are BJD 2455595--2455598, BJD 2455602--2455604.5, 
BJD 2455608--2455611, BJD 2455618--2455621, and these 
data is folded by the mean negative superhump period of 
these data, 0.0623106 d).  We also estimated the amplitude 
of negative superhumps and showed in the same figure. 

 The resultant $O-C$ diagrams are figures \ref{oc11}, figure \ref{oc12}. 
 The $O-C$ diagrams indicate that the period of negative 
superhump period gradually shortens as the next superoutburst 
approaches. Namely the derivative of negative superhump 
period $\dot{P}_{\rm{nSH}}$ is negative between the successive 
superoutburst.  The value of $\dot{P}_{\rm{nSH}}$ in 2011 
is $-1.10(6)\times 10^{-5}$ (supercycle 2011 S1), 
$-1.32(4) \times 10^{-5}$ (supercycle 2011 S2), $-1.04(11) \times 10^{-5}$ 
(supercycle 2011 S3). Meanwhile the value of $\dot{P}_{\rm{nSH}}$ 
in 2012 is $-5(2)\times 10^{-6}$ (supercycle 2012 S0), 
$-9.7(4) \times 10^{-6}$ (supercycle 2012 S1), $-7.7(6) \times 10^{-6}$ 
(supercycle 2012 S2). The absolute value of $\dot{P}_{\rm{nSH}}$ 
 in 2011 is larger than that in 2012.
 However, the figure of $O-C$ diagram shown in figure 
\ref{oc12} includes more complicated structures with 
shorter time-scales, or from the view of normal outburst
cycle. Namely, these $O-C$ curves are composed of 
multiple curves of concave-up shape although the general 
appearance was concave down in long time-scale, namely
from the view of supercycle.
 The third panel of figures \ref{oc11}, \ref{oc12} 
is denoted to the change of period during 
each supercycle. These values were calculated by PDM 
analysis for a 5-d interval.  Since the width is near the
interval of normal outburst, it is hard to detect clear
change of the period when interval of two normal outburst
is short.

 Since the period of the negative superhump is shorter than 
the orbital period, small period of negative superhumps 
corresponds to the large negative fractional superhump deficit
$\epsilon_{-}$. Thus this result implies that the absolute 
value of $\epsilon_{-}$ gradually increases as the next 
superoutburst approaches, and in shorter time-scale, the value 
of $\epsilon_{-}$ abruptly increases at the start of each normal 
outburst and gradually decreases until the next outburst starts.
The increase of $\epsilon_{-}$ in the longer time-scale is 
because that the abrupt increase in the rising stage is larger 
than the gradual decrease in quiescence.

\subsubsection{Lasso Period Analysis}

\begin{figure}
  \begin{center}
    \FigureFile(160mm,128mm){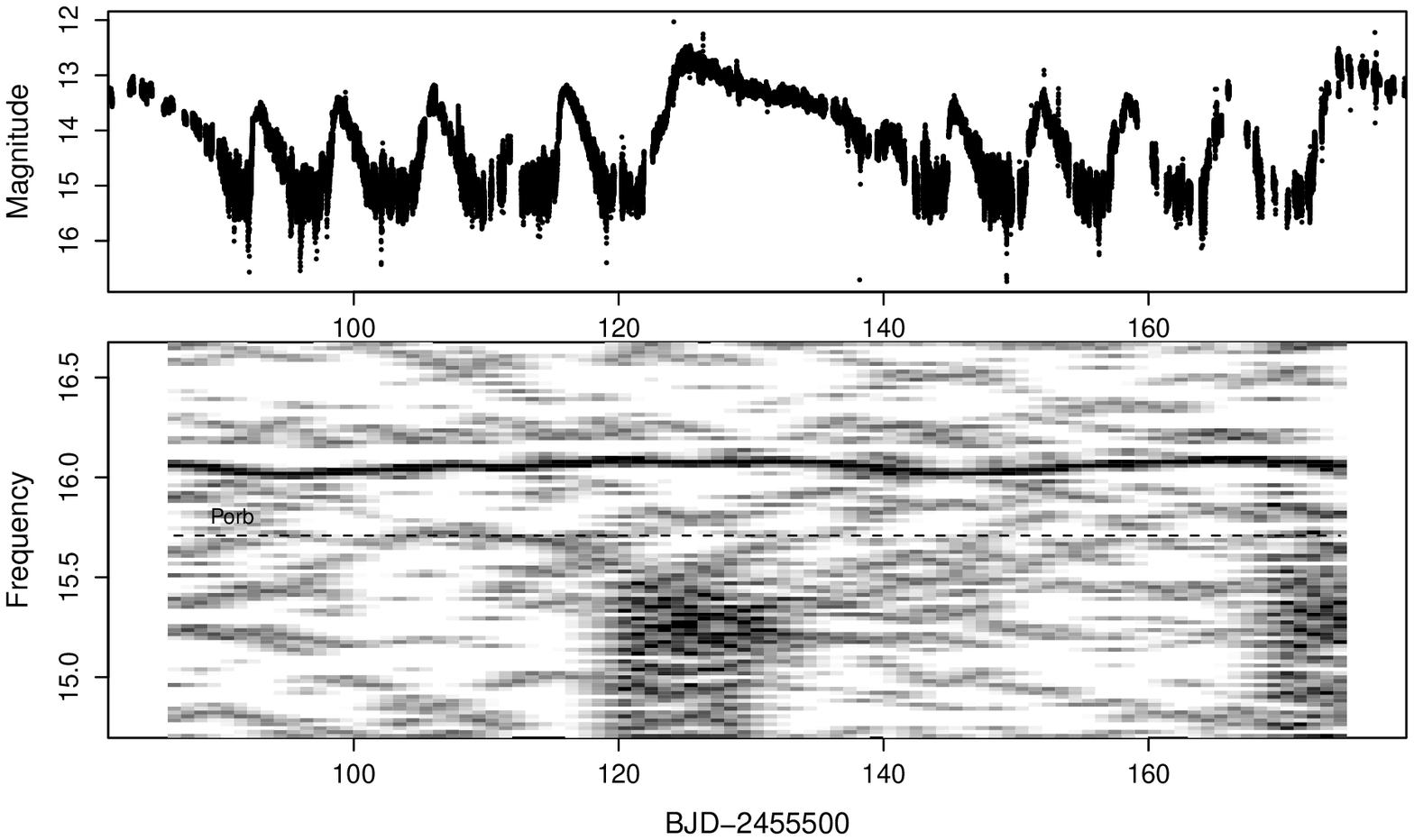}
    \FigureFile(160mm,128mm){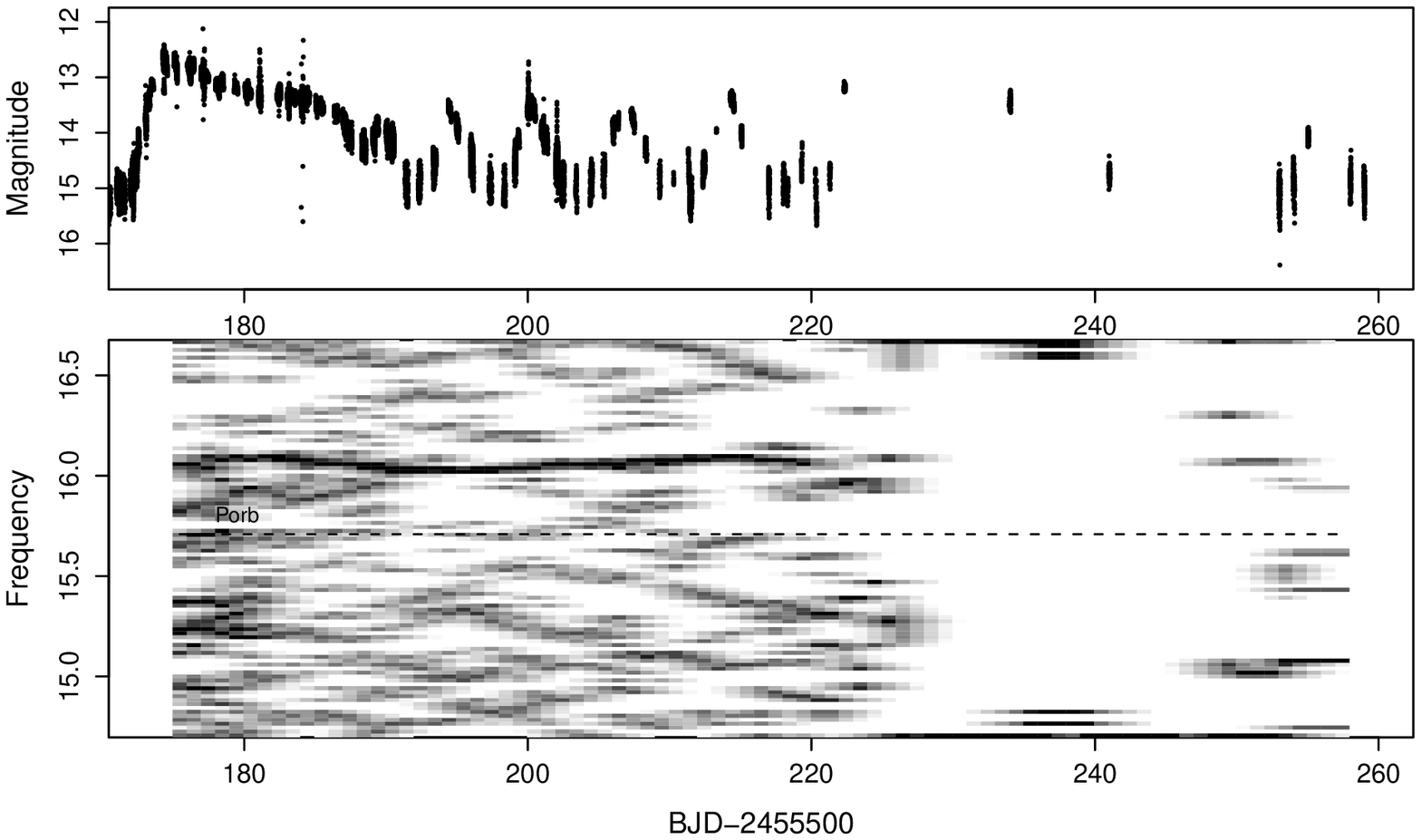}
  \end{center}
  \caption{Two dimensional period analysis of ER 
UMa using Lasso in 2011 season. For each two panels, 
the upper panel is the light curve and the lower panel
is the power spectrum. The width of the window is 10 d,
and the time step is 1 d. }
  \label{fig:lasso1}
\end{figure}

\begin{figure}
  \begin{center}
    \FigureFile(160mm,128mm){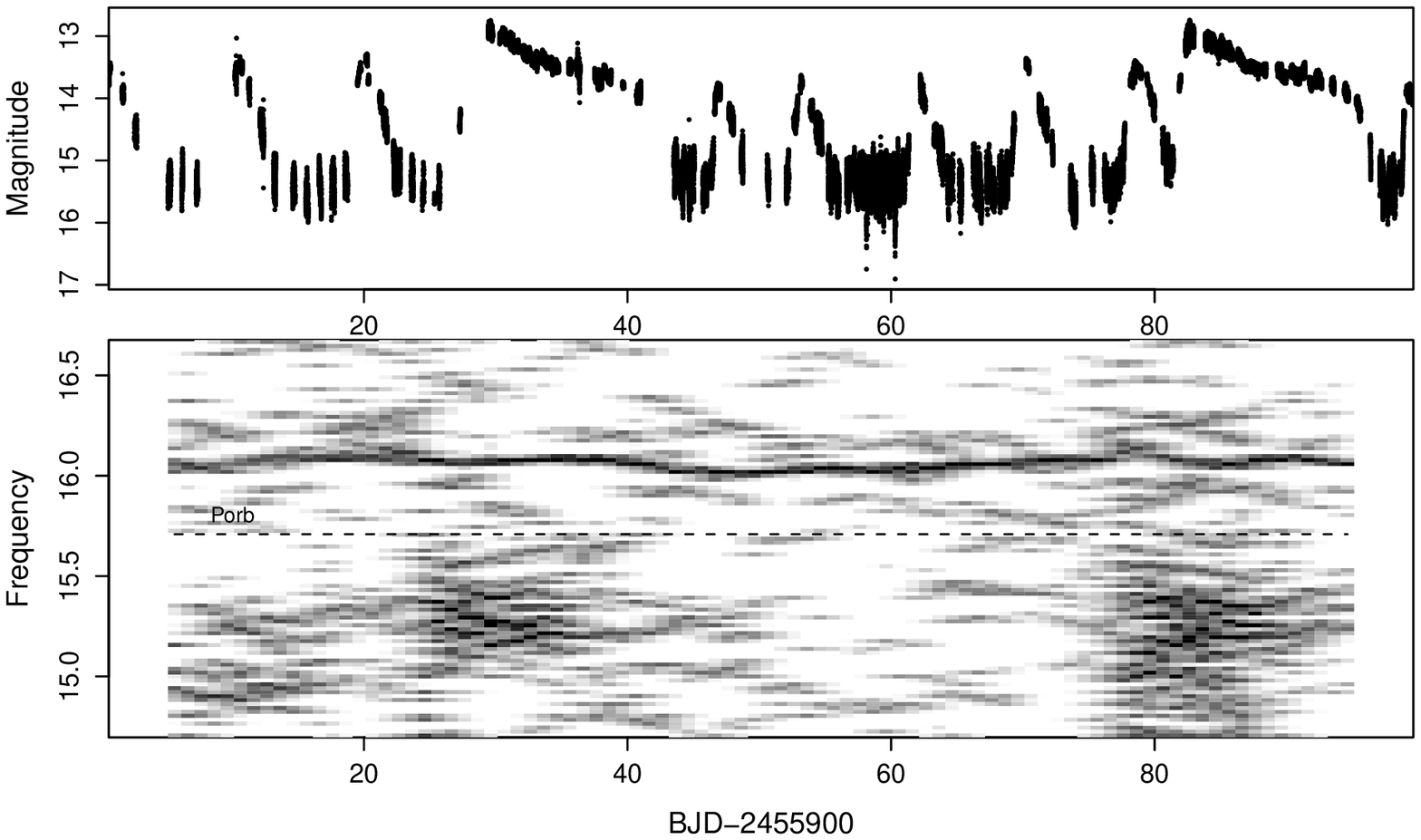}
    \FigureFile(160mm,128mm){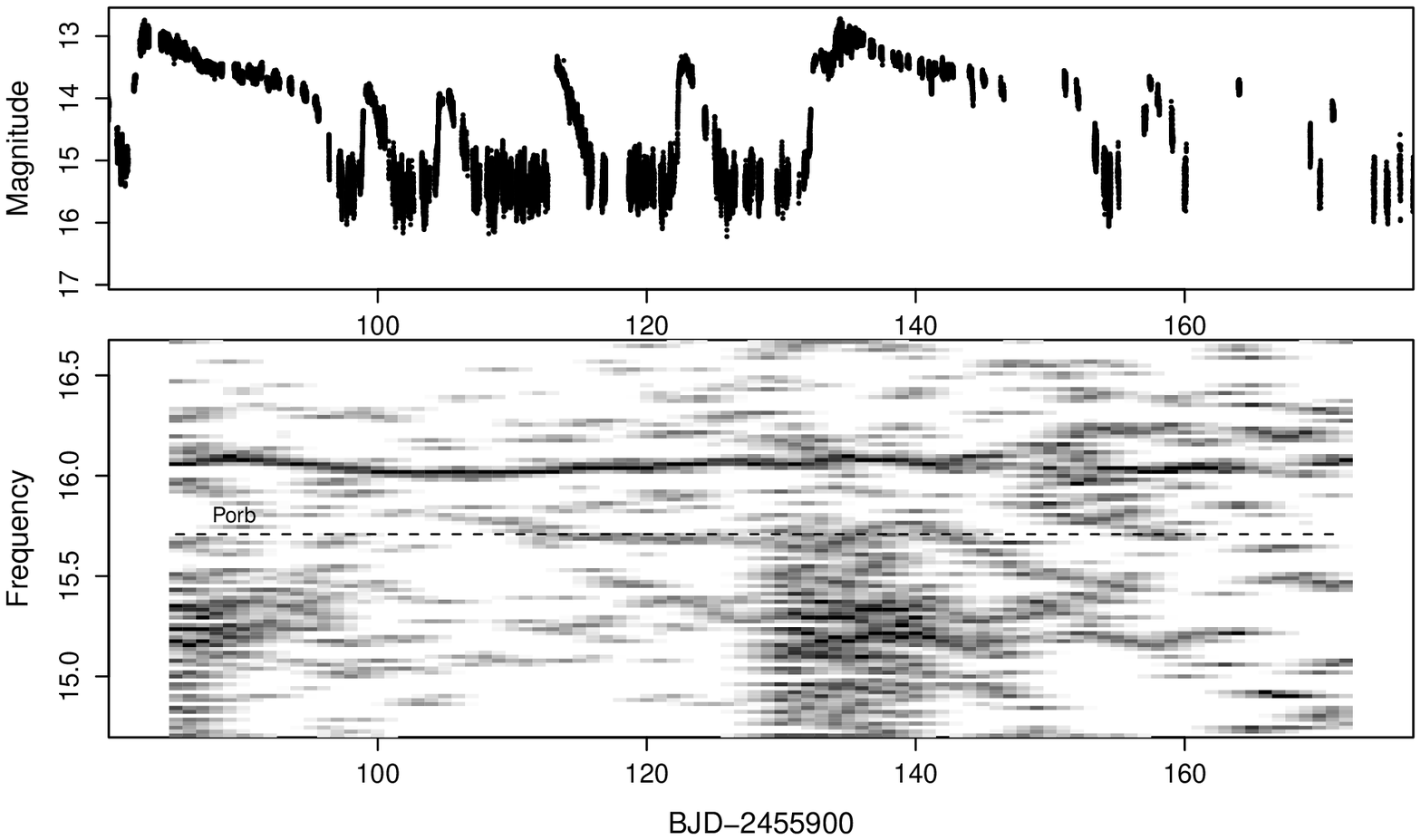}
  \end{center}
  \caption{Two dimensional period analysis of 
ER UMa using Lasso in 2012 season. For each two panels, 
the upper panel is the light curve and the lower panel
is the power spectrum. The width of the window is 10 d,
and the time step is 1 d. }
  \label{fig:lasso2}
\end{figure}

 Negative superhumps of ER UMa existed almost always 
during observations of 2011 and 2012. We made a 
detailed analysis for the frequency variations of 
negative superhumps.
 We computed two-dimensional power spectra of the 
light curve of ER UMa. We used locally weighted 
polynoinal regression (LOWESS: \cite{LOWESS}) to 
the observation data in order to remove trends 
resulting from outbursts with R software \footnotemark[2].
After that, we estimated the pulsed flux by multiplying 
the residual amplitudes and LOWESS-smoothed light curve 
converted to the flux scale. We used 10 d with of the moving
 window, and 1 d as the time step because the data is not as
contiguous as Kepler data.  Since the window length is
longer than the normal outburst cycle, the periodic change
of shorter time-scale is not well resolved. However, the
periodic variation of longer
time-scale is clearly seen.

 We performed a period analysis called least absolute 
shrinkage and selection operator (Lasso, \cite{lasso}),
 which was introduced to analysis of astronomical 
time-series data (\cite{kat12perlasso}, \cite{kat13j1922}). 
This method is very suitable to find peaks in power 
spectra and very strong method to analysis the rapid 
change of the period as in outbursting dwarf novae, 
because Lasso analysis has the advantage that peaks in 
power spectra are very sharp, and that it is less affected 
by uneven sampling data than Fourier analysis.

\footnotetext[2]{http://www.r-project.org/index.html}

 The resultant two-dimensional power spectra are shown 
in figures \ref{fig:lasso1}, \ref{fig:lasso2}.  These 
figures show that the clear signal of negative superhump
 signal was always detected during two seasons, except 
for the early stage of superoutburst and later phase of 
2011 season. Positive superhump signal was detected only 
in the early stage of superoutburst. The frequency of negative 
superhump changes during the supercycle. The frequency of negative 
superhump was smallest when the superoutburst ended and increases
toward the next superoutburst.

 Interestingly the Lasso diagram, especially in 2012 
show orbital modulation is detected. Although we tried 
to the period of orbital modulation, significant signal 
is not seen  because the signal was seen only partly.

 \citet{Pdot5} and \citet{osa13v1504cygv344lyrpaper3} showed 
that the variation of negative superhump
period in BK Lyn and ER UMa is not so large as that of ordinary 
SU UMa-type, such as V344 Lyr and V1504 Cyg and suggested this
is because of the interval of these objects is extremely
short.
 
\subsubsection{The Discussion about the Periodic Variation in Negative Superhumps}

 Both the $O-C$ diagram and the Lasso analysis show that the 
negative superhump period shortens as the next superoutburst 
approaches in long time-scale. However, in short time-scale,
 the period of negative superhumps tends to became longer 
in quiescence and an 
abrupt shortening occurs at the start of normal outbursts. 
By the combination of these 
two effects, the period of negative superhump period becomes 
shorter as a whole accompanied by smaller variations 
coinciding with normal outbursts outside the superoutburst 
stage. This change corresponds to the global form of the 
$O-C$ diagram. The period change estimated by the PDM analysis 
is also in good agreement with this result. 

 The theoretical relation implies that the increase of 
$\nu_{nSH}$ can be interpreted as the increase of $R_{\rm{d}}/A$. 
Thus the change of negative superhump period can be interpreted 
that the radius of the disk increases when the normal outburst
 is triggered and the accretion disk shrinks until the next 
normal outburst starts although the increase is larger than 
the decrease.
 This change of disk radius is similar that of V1504 Cyg shown in 
(\cite{osa13v344lyrv1504cyg}\cite{osa13v344lyrv1504cyg}). 

 The TTI model suggests that the increase of disk radius at 
the start of outburst because of the conservation of angular 
momentum and the increased viscosity \citep{osa89suuma}. After 
the outburst has finished, the disk radius shrinks until the 
next outburst starts. The negative superhump period becomes 
shorter as the next superoutburst approaches. Our result obeys 
this trend.

\subsection{Positive superhump}



\subsubsection{The Stage A Superhumps}

\begin{figure}
  \begin{center}
    \FigureFile(120mm,75mm){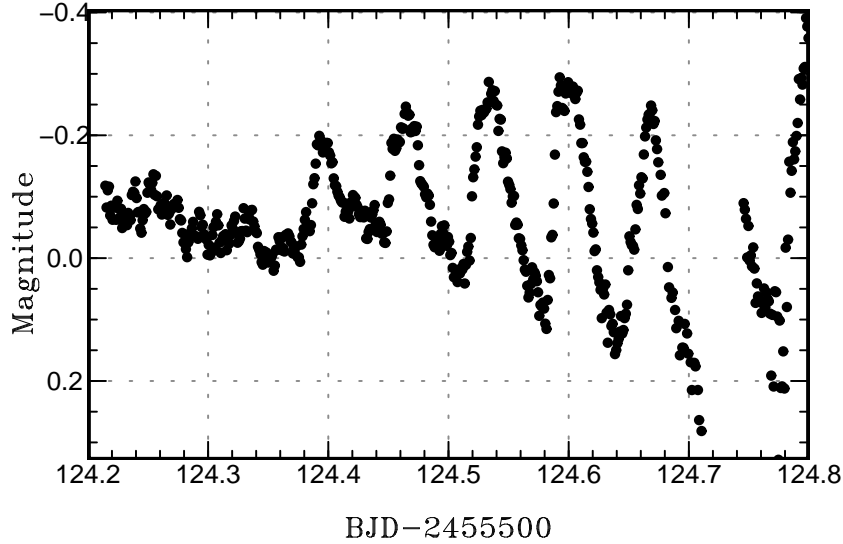}
  \end{center}
  \caption{Stage A superhump of 2011 S2. LOWESS fitting and
the subtraction of negative superhump signal is already taken.
 The observed data is binned to 0.001 d.}
 \label{stagea}
\end{figure}

\begin{figure}
  \begin{center}
    \FigureFile(72mm,108mm){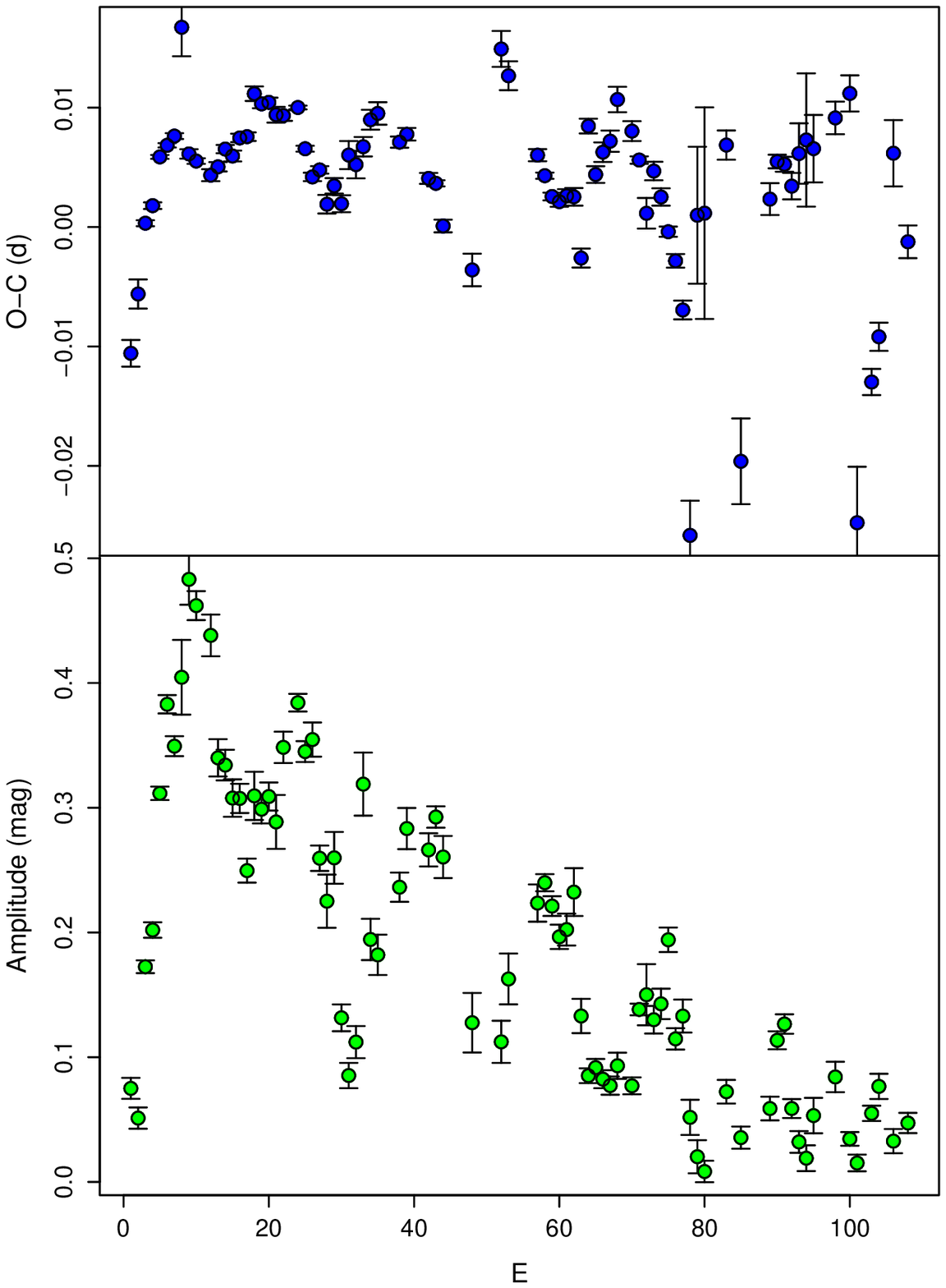}
    \FigureFile(72mm,108mm){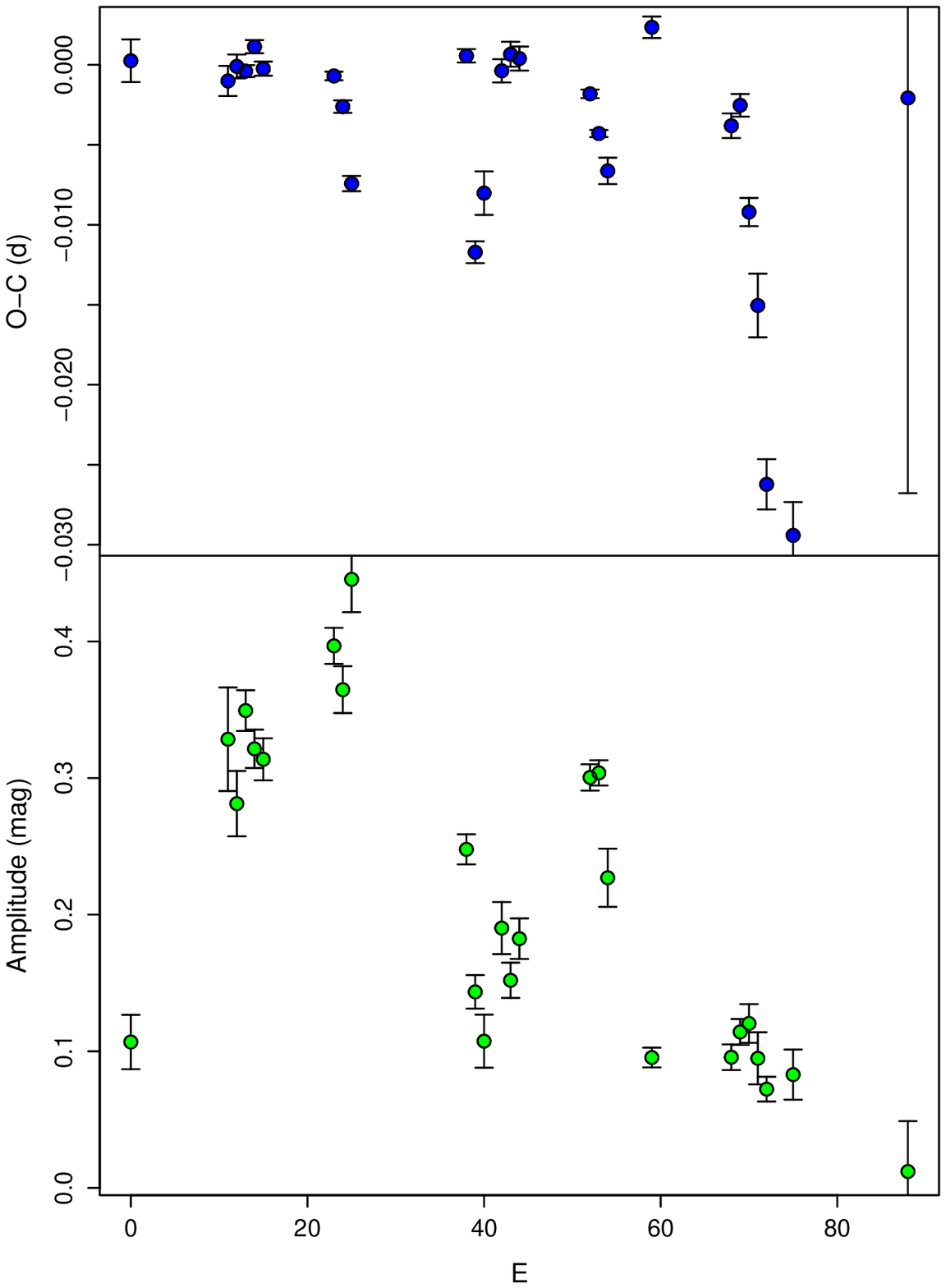}
  \end{center}
  \caption{$O-C$ diagrams of positive superhumps after 
subtracting negative superhumps of superoutbursts 2011 S2, 
S3. The $O-C$ values are against the equation of  
$2455624.392 + 0.065619 E$ (for 2011 S2) and $2455674.280 + 
0.065619 E$ (for 2012 S3).}
 \label{ocpSH11}
\end{figure}

\begin{figure}
     \FigureFile(72mm,108mm){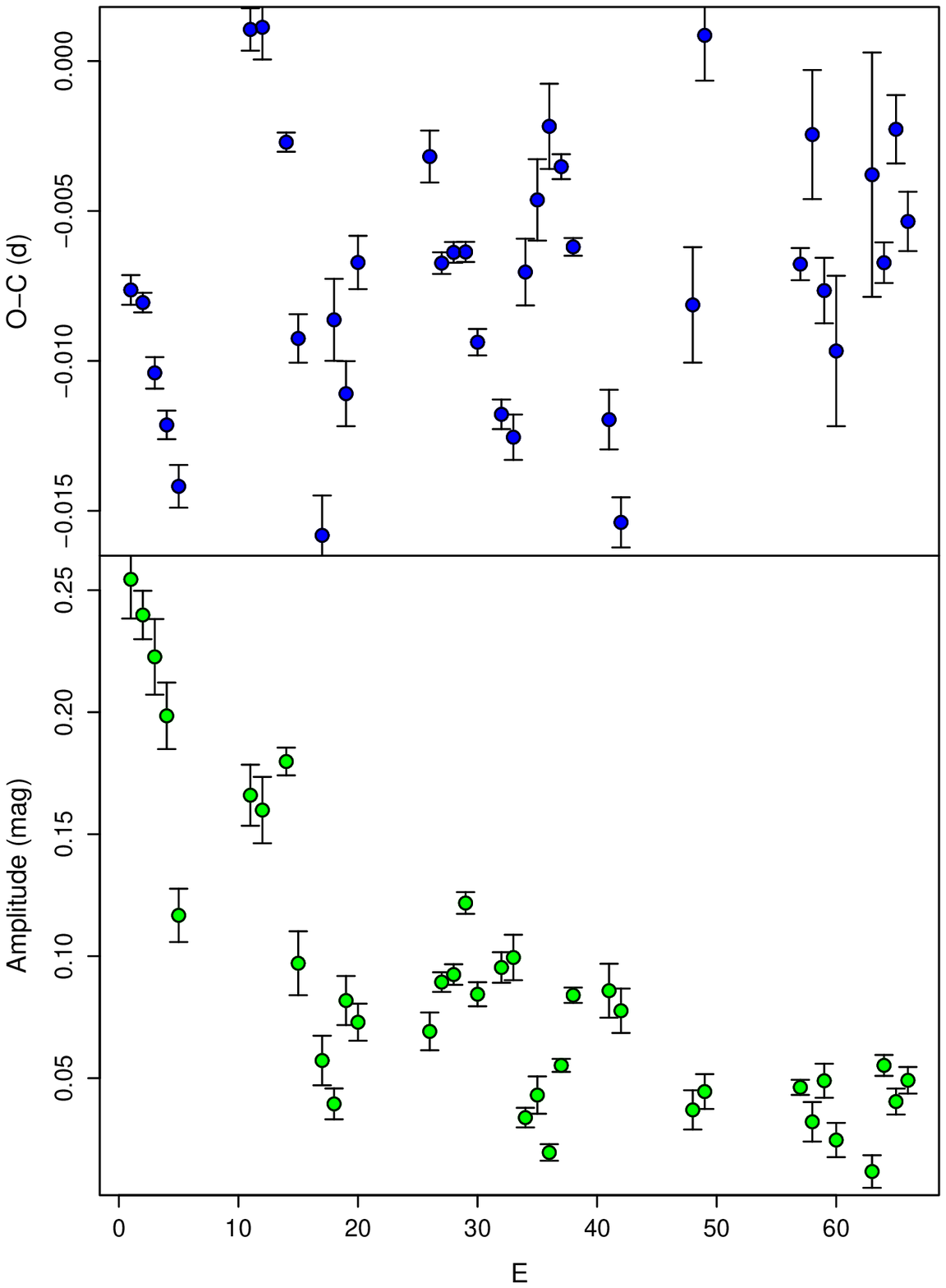}
    \FigureFile(72mm,108mm){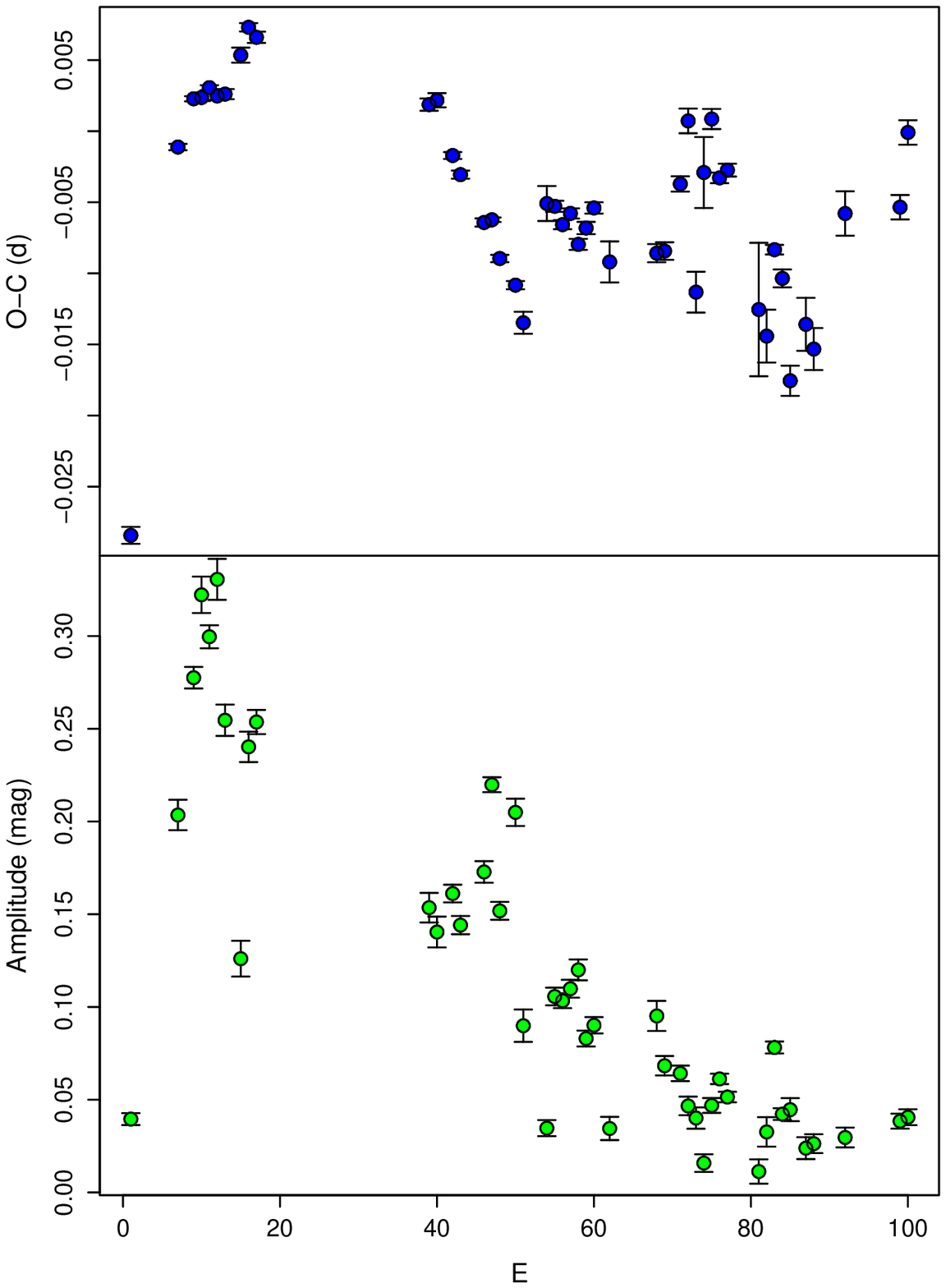}
    \FigureFile(72mm,108mm){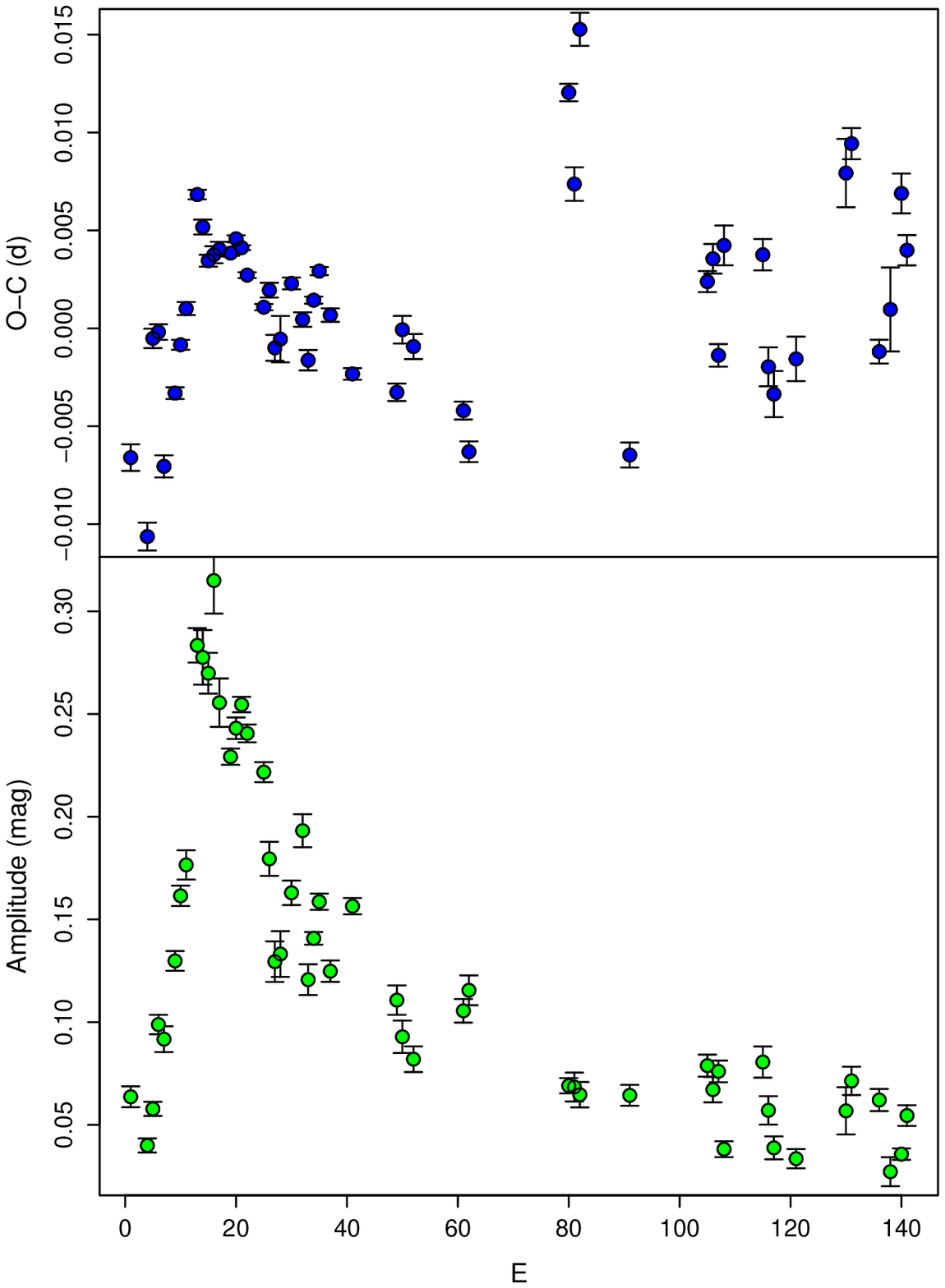}
  \caption{$O-C$ diagrams of positive superhumps after 
subtracting negative superhumps of 2012 S1 -- S3. The value 
is against the equation of  $2455924.094 + 0.065619E$ 
(for 2012 S1), $2455982.403 + 0.065619E$ (for 2012 S2), 
$2456034.104 + 0.065710E$ (for 2012 S3)}
 \label{ocpSH12}
\end{figure}

 ER UMa also shows positive superhumps.
 In recent researches (e.g. \cite{Pdot}), a superoutburst can
divided to three stages named stage A, B, and C from the change of superhump
period. Stage A corresponds to the evolving phase of superhump, when
the tidal instability is limited within 3:1 resonance radius.
After the superhumps after stage B,the eccentric wave spread to 
inner region of the disk and the pressure effect appears
\citep{osa13v344lyrv1504cyg}. Thus stage A superhump period gives us 
the mass ratio $q$ of the system \citep{kat13qfromstageA}.
It is very useful to detect stage A superhump and estimate the period.

 \citet{kat13qfromstageA} suggested an explanation why
stage A superhumps are difficult to detect in ER UMa-type
dwarf novae.  In ER UMa-type dwarf novae, the superoutburst 
is not necessarily triggered by a normal outburst but
by the eccentric instability 
[called Case C outburst \citet{osa03DNoutburst}]. 
In such case, the pressure effect has already been strong 
at the start of a superoutburst and the method to estimate 
$q$ with stage A superhump period may not be applicable.
 However, in superoutbursts during our observations, positive
superhumps were triggered by normal outburst as well as other 
usual SU UMa-type objects. Therefore the existence of the
stage A superhump is expected.

  As 
seen in figures \ref{fig:lasso1} and \ref{fig:lasso2}, 
negative and positive superhumps co-exist during the 
superoutburst. It is supposed that positive superhump is 
caused by prograde precession of the elliptical disk and negative
 superhumps are caused by retrograde precession of the tilted 
disk. The co-existence of positive superhump and negative 
superhump suggests that the disk is eccentric and tilted at 
the same time. 

 This co-existence of negative and positive superhumps is
a problem to estimate the maximum timings of positive superhumps.
We have to subtract the
variations of negative superhumps.

  We adopted the averaged light curve 
of negative superhump used for the subtraction.
First we subtracted from
the original light curve translated to the flux scale.
After that, the averaged light 
curve was formed data subset during one beat cycle. With
this averaged light curve, 
After the subtraction, the scale was translated to the magnitude
scale again. The subtracted light curve is figure \ref{stagea}.

After subtraction of negative superhumps, The $O-C$ 
diagrams of five superoutbursts are figures \ref{ocpSH11},
\ref{ocpSH12}. 
In 2011 S1, negative superhump was dominant 
at the start of time-resolved observation, thus $O-C$ curve
of positive superhump could not be drawn. 

 Among five superoutbursts, stage A superhumps were
detected in three superoutbursts (2011 S2, 2012 S2, and 2012 S3).
These detection were based on the longer superhump period and the 
increase of the amplitude of superhumps in the earliest
stage of the superoutburst.
For instance,  the amplitude of the positive superhumps evolved to
0.25 mag until $E = 10$ in 2011 S2 (figure \ref{stagea}). In 2011 S3 and 
2012 S1, it was difficult to estimate stage A superhump 
period because of the lack of observation.
 After the amplitude of the positive superhumps reached 
the maximum, the amplitude of positive superhump became 
gradually smaller. These can be regarded as stage B 
superhumps. The perfect 
subtraction of negative superhump, especially in the later 
stage, is difficult, however. The profile of superhumps
of ER UMa during superoutburst does no seem to be simple superposition
of positive and negative superhumps.

 We then obtained the periods of stage A superhumps.
 For the data of 2011 S2, A PDM 
analysis yielded a stage A superhump period of 0.06604(9) d. 
 Similarly the stage A superhump period is estimated as 
0.06570(2) d in 2012S2 and 0.06624(4) d in 2012 S3.
 With this data, the estimated $q$ is 0.100(6) by the method
of \citet{kat13qfromstageA}. Data of 
other superoutbursts show somewhat different value, 0.088
 (2012 S2) and 0.114 (2012 S3). Using an average of these 
values, we adopted $q$ of 
ER UMa to be 0.100(15). 

\subsubsection{System Property and Evolutionary State}

\begin{figure}
   \begin{center}\label{evol}
     \FigureFile(120mm,90mm){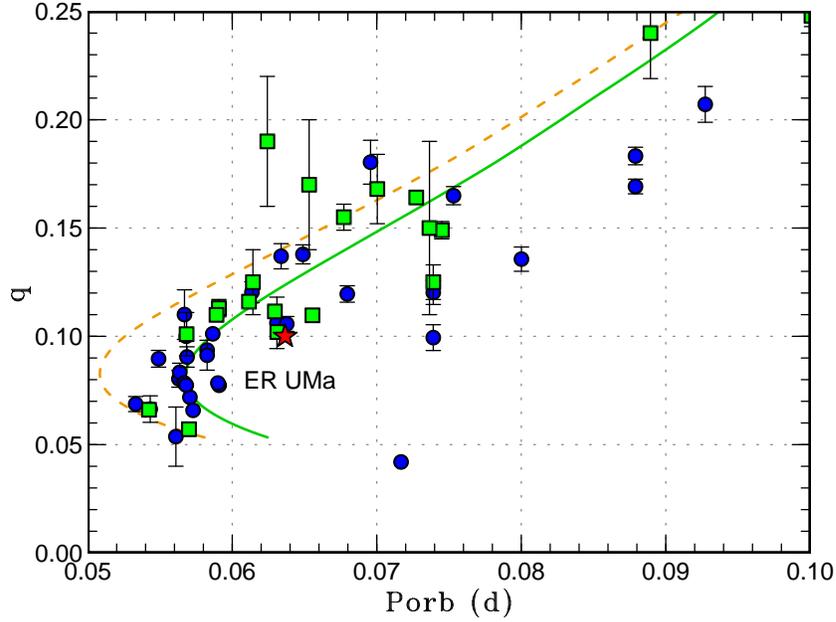}
   \end{center}
   \caption{Diagram of $q$ versus $P_{orb}$. The data are derived
from \citet{kat13qfromstageA}The filled circles represent $q$
estimated from stage A superhumps. filled squares represent
$q$ measured by eclipses.The filled star represents ER UMa. The
dash curved line and solid curved line represent evolutionary track
of the standard evolutional theory and that of the modified evolutional
theory \citep{kni11CVdonor}
}
 \end{figure}
 The estimated value of $q$, 0.100(15) suggests that ER UMa is
on the standard evolutionary track in \citet{kni11CVdonor} 
since the orbital period is 0.06366 d. Our results indicate
that these is no evidence that ER UMa is in the evolutional stage 
different from ordinary CVs although $\dot{M}$ of ER UMa is 
much higher than other SU UMa-type dwarf novae with
similar orbital periods.

\citet{hel01eruma} suggested that the unusual behavior of
ER UMa-type (rapid reccurence of normal outbursts)
or WZ Sge-type objects (rebrightenings) may be explained if 
these objects have extremely low $q$ (i.e. near 
the period minimum or period bouncers) and 
the thermal and tidal instabilities are decoupled due to 
the weak tidal force. 
Our present result indicates that at least ER UMa itself
is not the case.
We consider that there is no necessity to consider 
decoupling of the thermal and tidal instabilities
for ER UMa as shown by \citet{osa95eruma}, in which
the behavior of ER UMa can be reproduced by increasing
the $\dot{M}$, while there remains a possibility
for RZ LMi \citep{osa95rzlmi}.  Determination of
the orbital period and detection of stage A superhumps
for RZ LMi and DI UMa are desired to solve this problem. 

\subsubsection{The relation between Positive and Negative Superhumps}
\begin{figure}
  \begin{center}
    \FigureFile(75mm,75mm){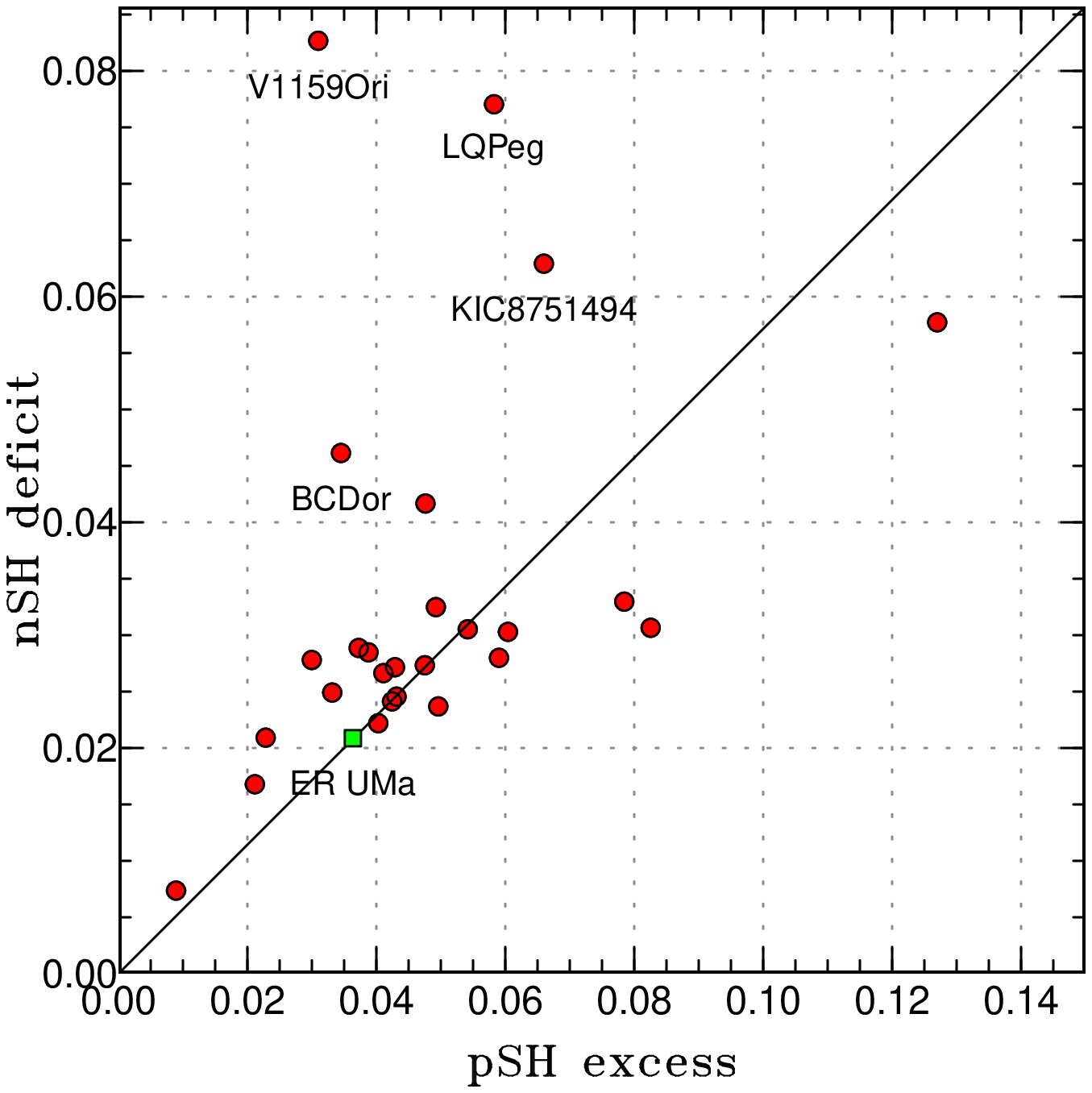}
    \FigureFile(75mm,75mm){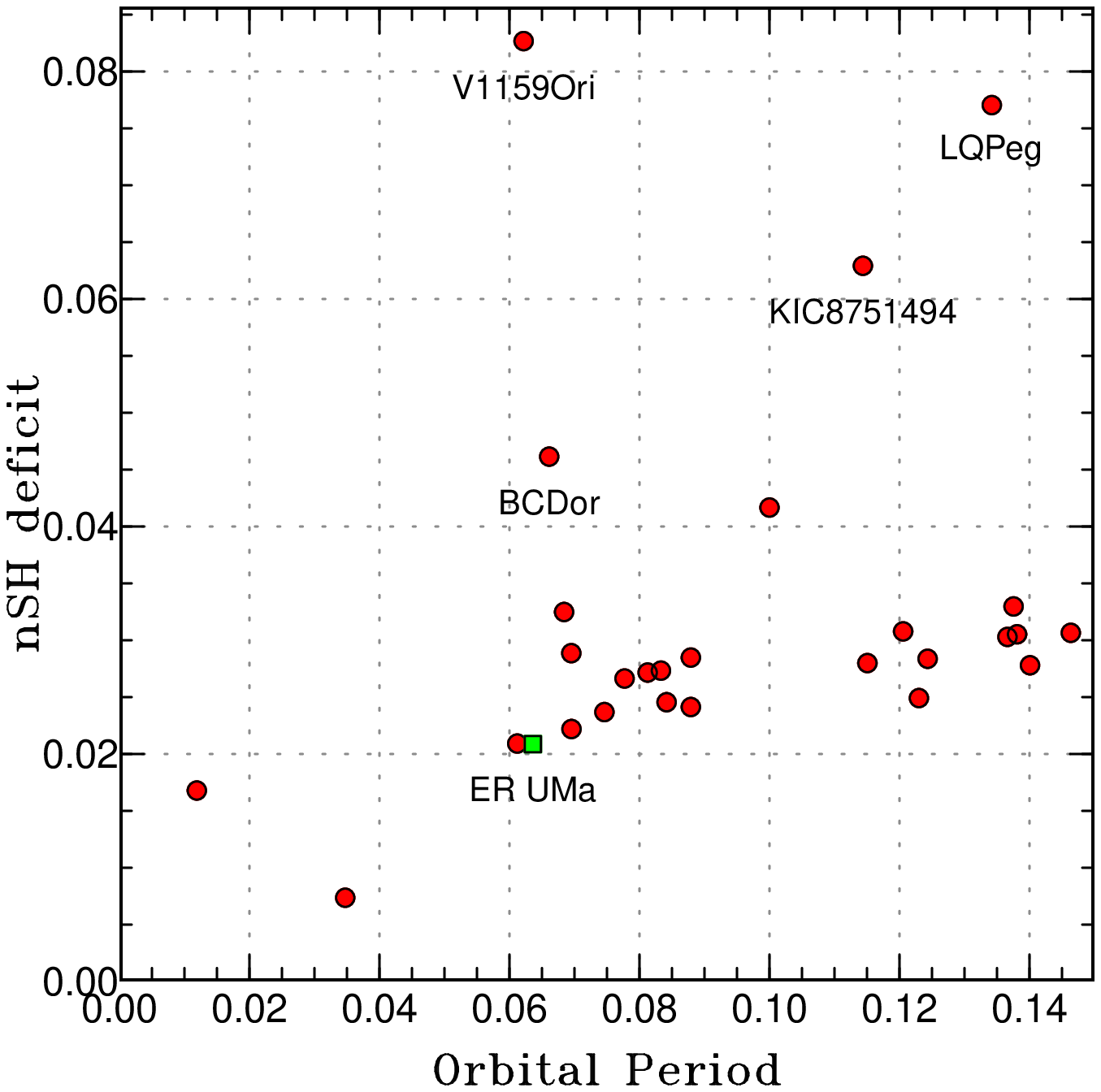}
  \end{center}
  \caption{$\epsilon_{+}$ vs. $\epsilon_{-}$ diagram. 
The solid line implies that the theoretical predicted 

relation in the absence of the pressure effect. 
Reference: V1159 Ori 
\citep{pat95v1159ori}, AM CVn (\cite{ski99amcvn}, 
\cite{pat98evolution}, \cite{pat99SH}), PX And 
\citep{sta02pxand}, TV Col \citep{ret03tvcol}, BF Ara 
(\cite{kat03bfara}, \cite{ole07bfara}),
V1405 Aql (\cite{cho01v1405aql} \cite{ret02v1405aqlproc}) 
AH Men \citep{pat95ahmen},
IR Gem \citep{fu04irgem},  V503 Cyg \citep{har95v503cyg},
TT Ari (\cite{ski98ttari},\cite{and99ttari},\cite{wu02ttari}), 
V603 Aql \citep{pat97v603aql}, 
RR Cha \citep{wou02rscarv365carv436carapcrurrchabioricmphev522sgr},
V344 Lyr \cite{sti10v344lyr} \citep{osa13v344lyrv1504cyg},
V1504 Cyg (\cite{osa13v344lyrv1504cyg}, \cite{osa13v1504cygKepler}),
QU Aqr (\cite{ole09j2100}, \cite{tra05j2100}),
BC Dor \citep{wou05CVphot},
DW UMa (\cite{sta04dwuma}, \cite{pat05SH}),
V1974 Cyg \citep{ole02v1974cyg},
KIC 8751494 \citep{kat13j1924},
CSS 091121:033232$+$020439 \citep{wou12SDSSCRTSCVs},
KIC 7524178 \citep{kat13j1922}
}
\label{nshpsh}
\end{figure}

 The left panel of figure \ref{nshpsh} shows the relation 
between the negative superhump period and the positive 
superhump period for systems which show both superhumps. 
Theoretically, the ratio the negative superhump period to 
the positive superhump period is 4/7 when the pressure 
effect do not effect. Most of that of systems obeys this 
relation. Among systems deviating from the theoretical 
relation, KIC 8751494 can be explained by the pressure 
effect \citet{kat13j1924}. The negative superhump of 
V1159 Ori may be not true negative superhumps, but 
``impulsive negative superhumps'' \citep{osa13v344lyrv1504cyg} 
\footnotemark[3].

\footnotetext[3] {This phenomenon was also described in 
\citet{woo11v344lyr}.}

 Our value of ER UMa is especially consistent with this 
relation. Although \citet{gao99erumaSH} is far shorter 
than our value and theoretical prediction, this may also 
be an ``impulsive negative superhump''.
The right panel of figure \ref{nshpsh} shows the relation 
between the orbital period and the negative superhump 
deficit.

\subsection{The Transition from Negative Superhump to Positive Superhump}

\begin{figure}
  \begin{center}
    \FigureFile(70mm,98mm){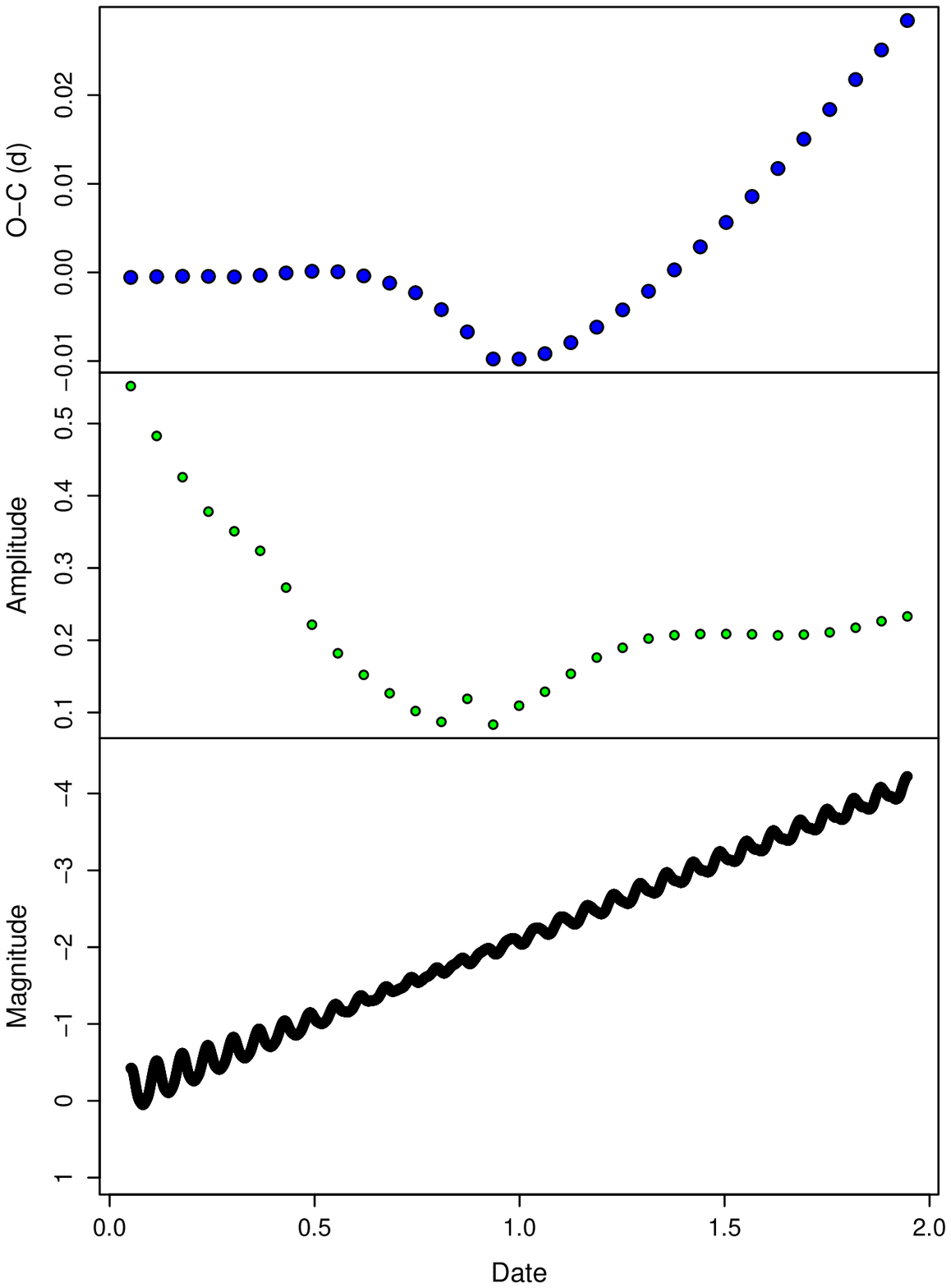}
    \FigureFile(70mm,98mm){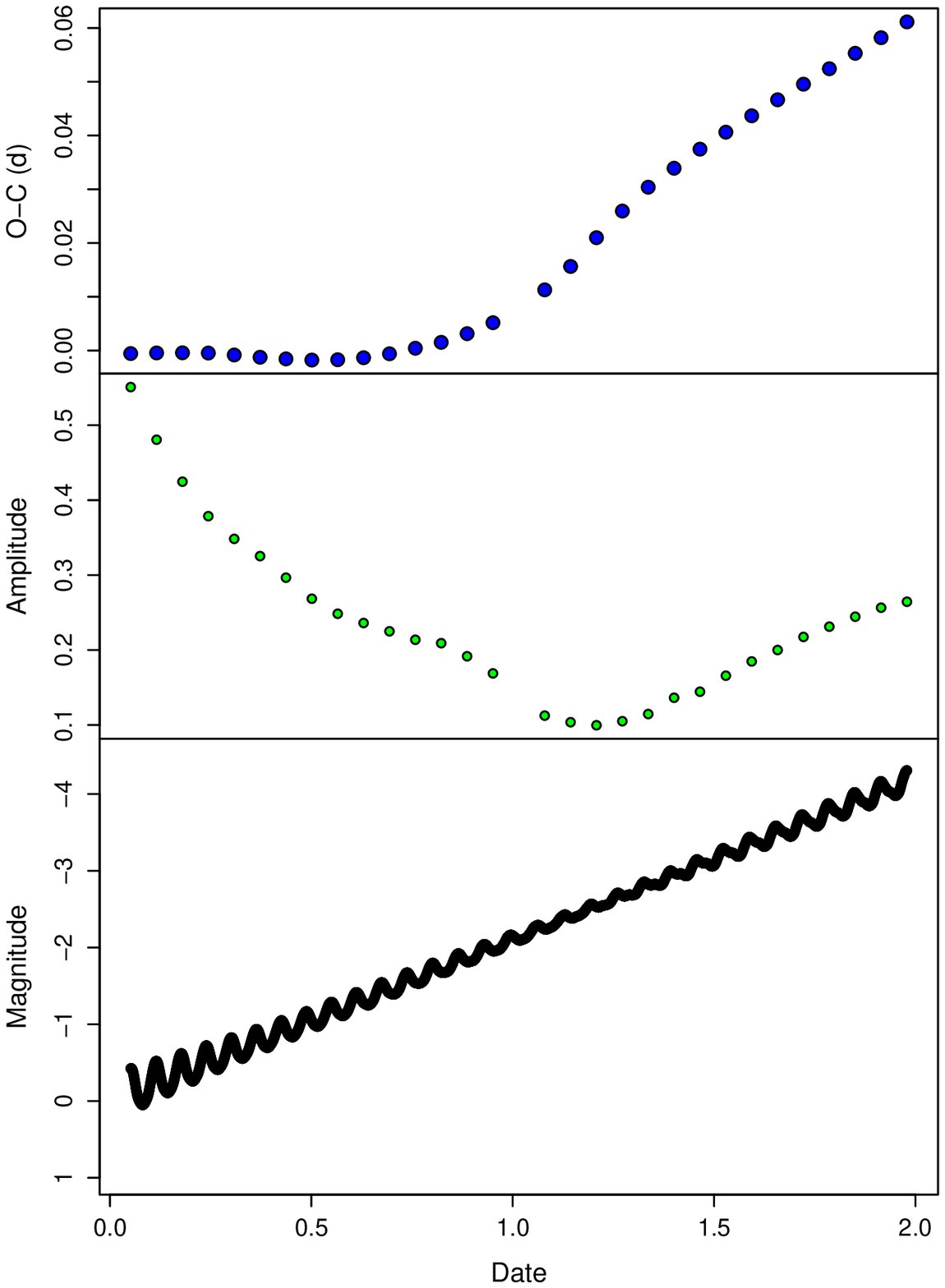}
  \end{center}
  \caption{Model calculation of the superimposition 
of negative and positive superhumps. In the left panel, 
the positive superhump starts at phase 0 when the phase 
of the negative superhump is also 0. In the right panel, 
the positive superhump starts at phase 0 when the phase 
of the negative superhump is 0.5. For each case, the 
variation of amplitude of hump and $O-C$ diagram at 
the transition stage from positive to negative 
superhump is shown.}
  \label{risingsimu}
\end{figure}

\begin{figure}
\begin{center}
 \FigureFile(60mm,70mm){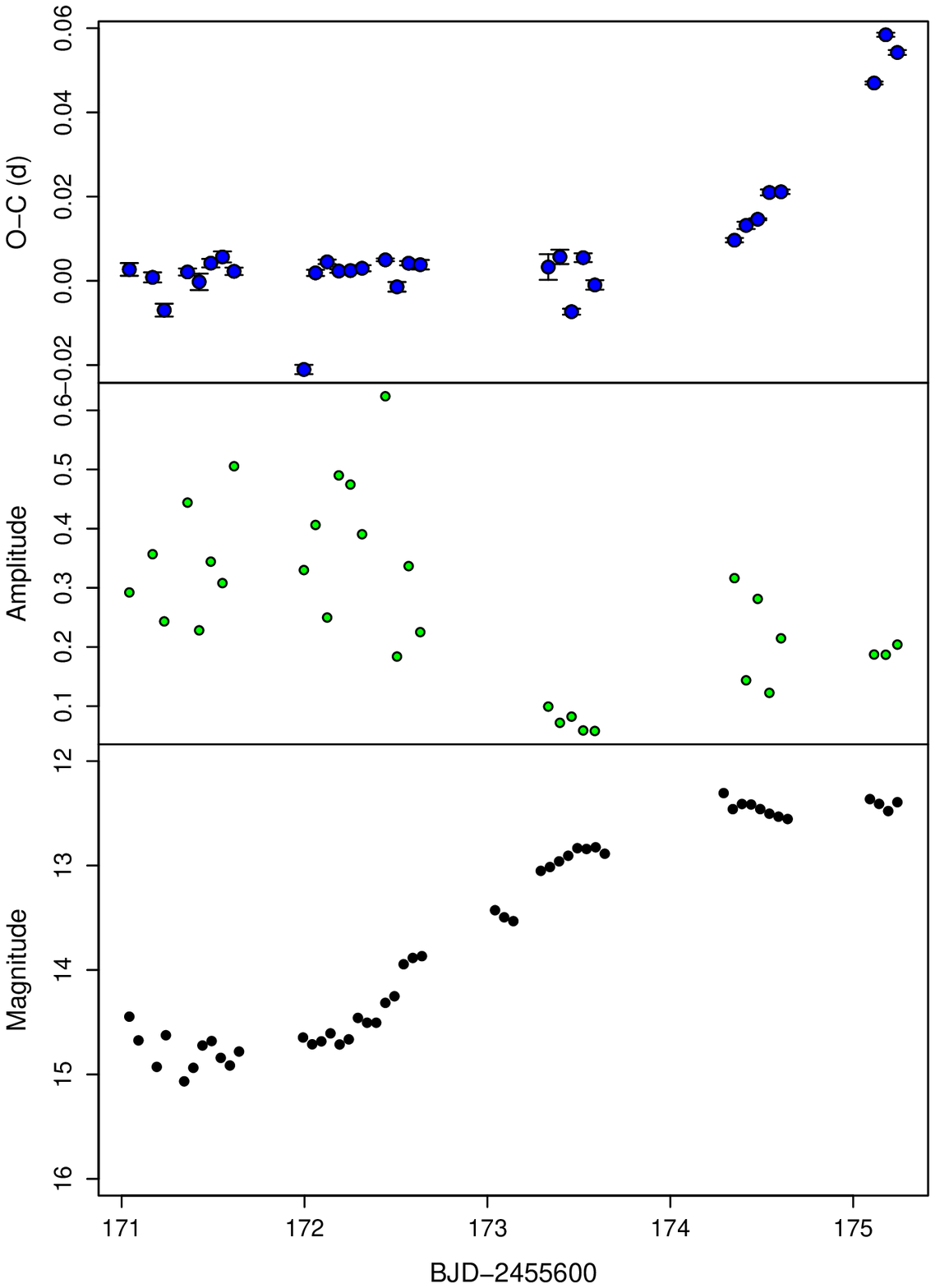}
\FigureFile(60mm,70mm){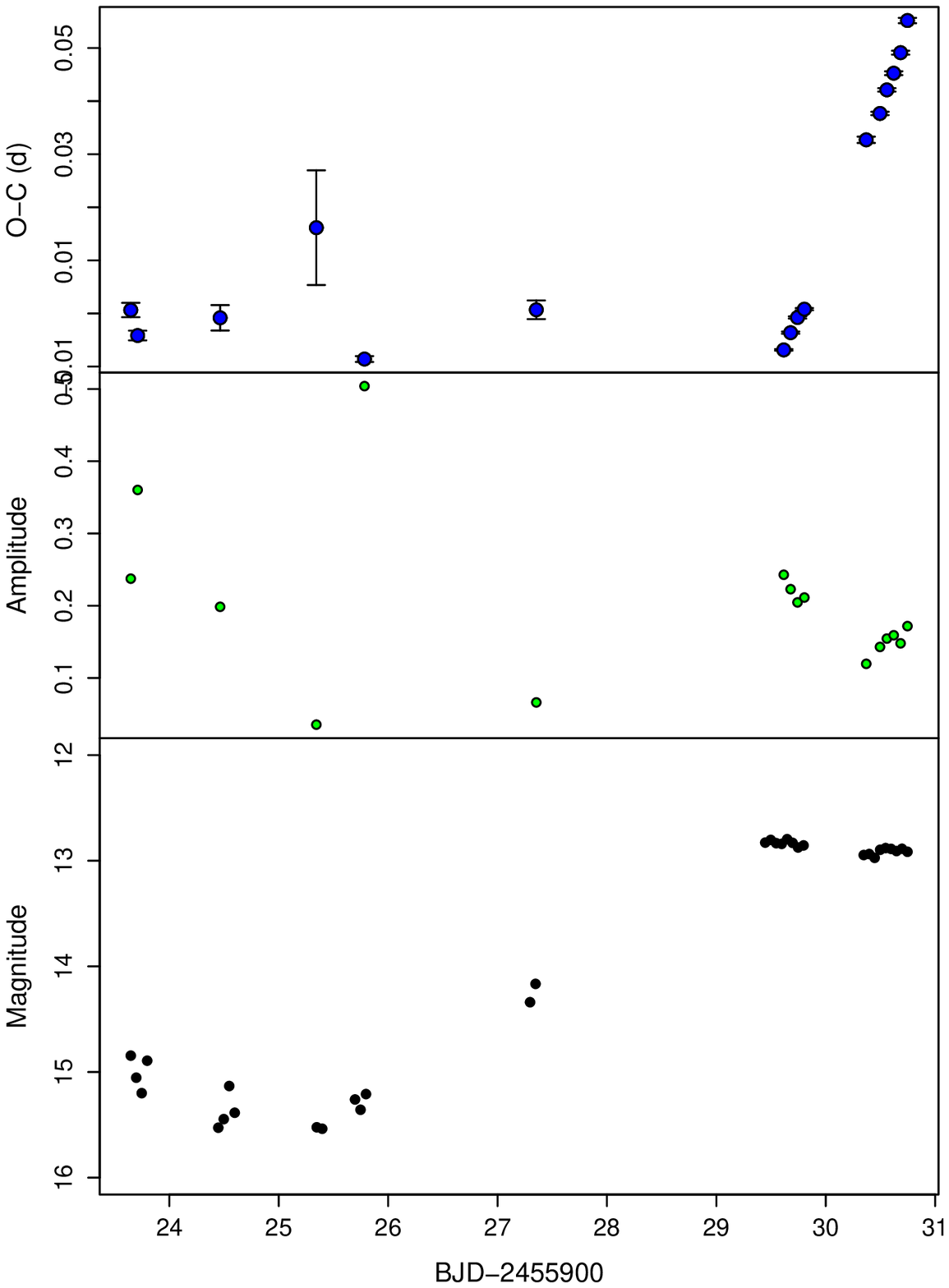}
\FigureFile(60mm,70mm){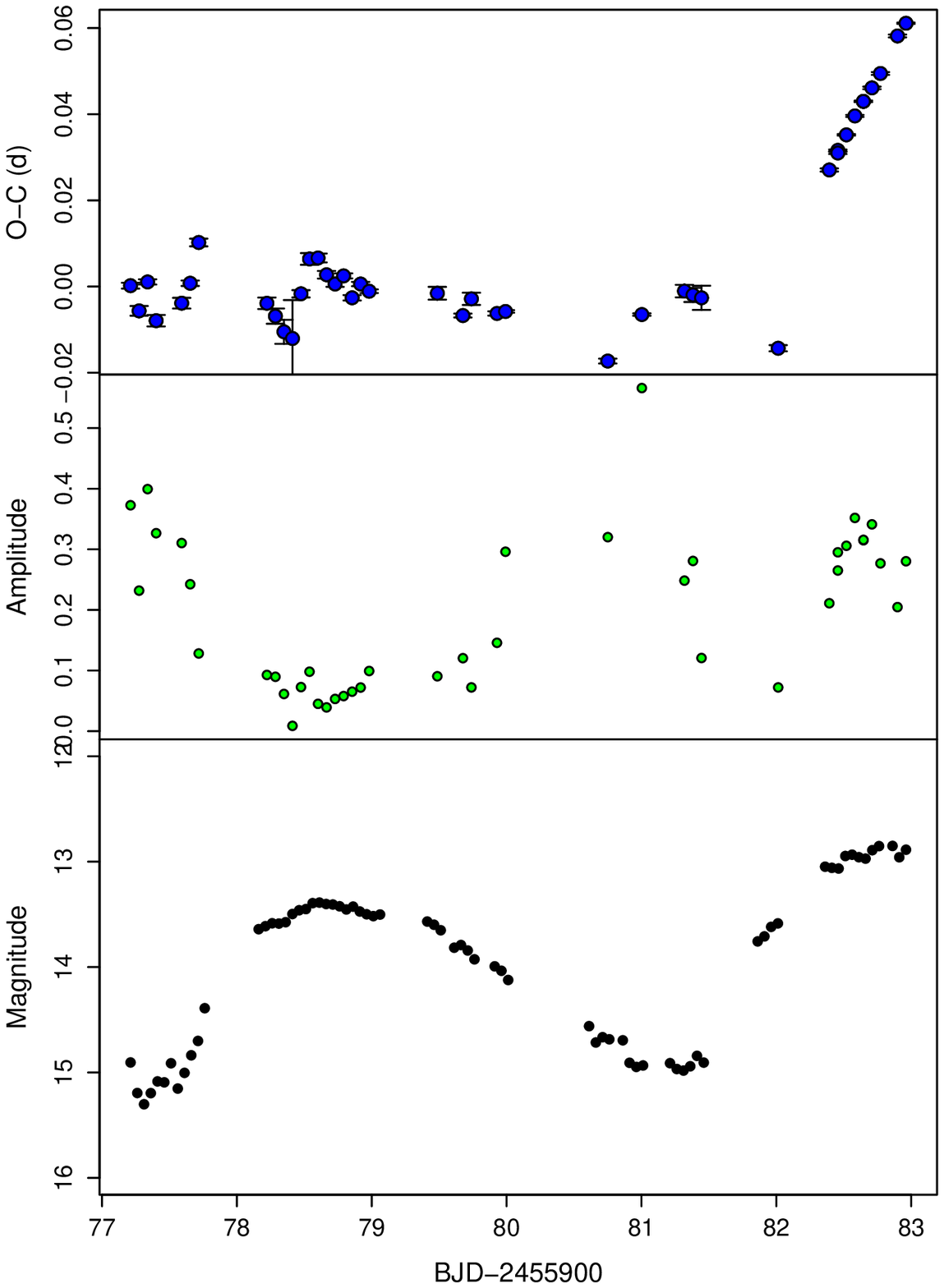}
\FigureFile(60mm,70mm){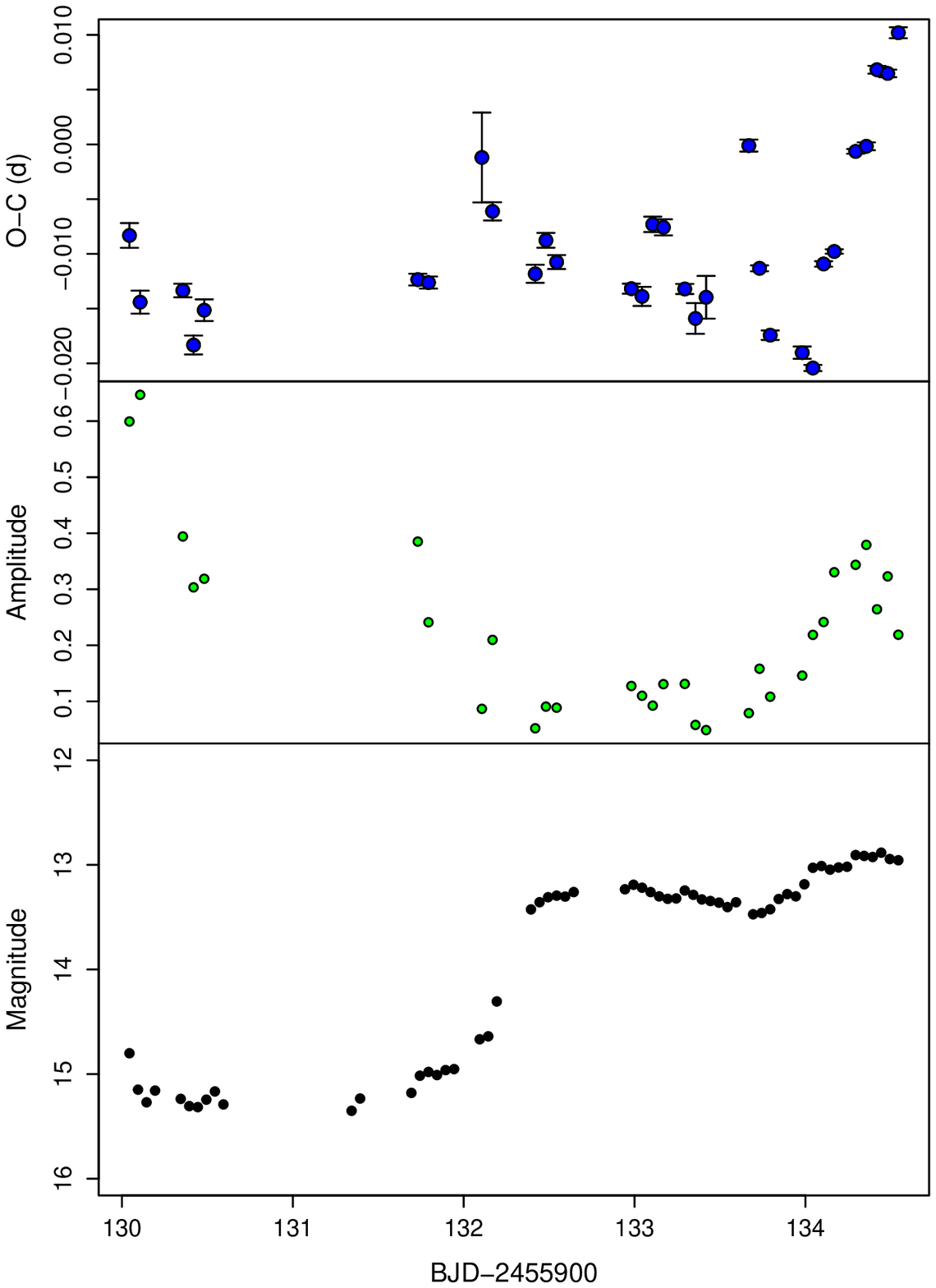}
  \end{center}
  \caption{$O-C$ diagrams, amplitude variation, and 
the light curve (0.01 d-binned) during the rising stage of 
superoutbursts. The upperleft panel is the  diagram of the 
superoutburst 2011S3. The values of $O-C$ diagram is against 
the equation of 2455671.039$+$0.0622$E$. The upperright panel 
is the  diagram of the superoutburst 2012S1. The values of 
$O-C$ diagram is against the equation of 2455923.646$+$0.0622$E$. 
The bottomleft panel is the diagram of the superoutburst 2012S2.
 The values of $O-C$ diagram is against the equation of 
2456077.211$+$ 0.0622$E$. The bottomright panel is the diagram 
of the superoutburst 2012S3. The values of $O-C$ diagram is 
against the equation of 2456077.211$+$ 0.0622$E$. For 2011S1, 
see fig 3 of Paper I. These diagrams suggests 
245630.052$+$0.0622$E$}
  \label{risingsoc}
\end{figure}

In paper I, we reported that the the maximum timings 
of negative and positive superhumps developed continuously 
and there was no phase shift between them. Paper I 
suggested that it implies the source of negative and positive 
superhumps is the same. 
The similar trend was seen also in other rising stages.
  However, this is unclear because the amplitude of superhumps
in transition stage was small.
 We tested this suggestion in model calculations. 

 We assumed the positive superhump develops in the rising stage 
of the superoutburst and the constant amplitude in flux of 
negative superhump, 0.7 mag in quiescence. The rising rate 
of mean magnitude is 2.1 mag / d.
 Positive superhumps to start 
the development four cycles after the rising starts. After positive
superhumps appearing, the amplitude
of positive develops at the speed of 0.15 mag /d. For the 
profile of negative and positive superhumps, the template 
profiles used for non-linear fitting were adopted.

 The resultant diagrams are shown in figure \ref{risingsimu}.
 In the upper diagram, the phases of negative and positive 
superhump are continuous when the positive superhumps start
 to develop. In the lower diagram, the phases of negative 
and positive superhump are different by 0.5. The both diagrams 
both show the decrease of the amplitude of variation when 
the positive superhumps begin to develop. Furthermore, the 
negative superhump phase evolve into  the positive superhump 
smoothly in the both $O-C$ diagrams. 
 In the upper panel of figure \ref{risingsimu}, 
$O-C$ variation does not show small a smooth transition 
from negative superhump to positive superhump, but shows 
more complex structure. This is caused by the superimposition
 of the maxima of negative and positive superhump. 
However, since the amplitude of positive 
superhump is small, the shift of $O-C$ value is small.

The $O-C$ diagrams of the rising stage of each superoutburst 
is shown in figure \ref{risingsoc}. These $O-C$ diagrams 
imply that the negative and positive superhumps are continuous
 without phase shift as in the lower case of figure 
\ref{risingsimu}. This result suggests that the position where 
the positive superhump is not randomly excited in relation 
to negative superhumps.

This result would not be expected if the negative superhumps
arise from the varying release of the potential energy
on a tilted disk and the phase of the tilt is random
to the observer when positive superhumps start to grow.
Either our understanding of the origin of negative
superhumps may be insufficient or the statistics may not
sufficient to prove whether the positive superhump is 
randomly excited in relation to negative superhumps.
Further systematic observations of the rising stage
of superoutbursts are required.

\section{Conclusion}
\label{conchap}

 We observed the SU UMa-type dwarf nova ER UMa in the years 2011 and
2012 by VSNET worldwide campaign and we obtained data of 307 nights
spaning in 2011 -- 2012. We 
detected  persistent negative superhumps during the superoutbursts as 
well as during normal outbursts and quiescence  in both
seasons. 
We analyzed these data 
and obtained the following are our major findings:

 (1) We succeeded in covering three supercycles of this star in
2011 and also three supercycles in 2012. The star showed persisted
superhumps in both seasons.

 (2) We analyzed periodic variations of negative superhumps in a
supercycle of this star by  $O-C$ analysis, PDM analysis, 
and Lasso analysis.  We have found from these analyses that
the period of negative superhumps between two 
successive superoutbursts decreases secularly from the
end of a superoutburst to the next superoutburst in a longer time-scale
of a supercycle. Superimposed on this long time-scale trend,  a 
shorter time-scale variation in the negative superhump period occurred
within a normal outburst cycle in a sense that the period was shortened
when a normal outburst occurred and it became
longer during quiescence.  When 
the next supercycle started, the period of negative superhumps
returned to the value when the previous supercycle started. 
 Thus it varied cyclically with the same supercycle period as that of
the light curve.
This variation was indicates as a result of the variation in
the disk radius and the disk radius was the smallest at the end of a
superoutburst. The variation in the disk radius  inferred by our
analysis of ER UMa is 
in agreement with the predicted variation of the disk radius by 
the TTI theory.

(3) The rising rate of maximum of the normal outburst varied during 
one supercycle. The rising rate  of the first normal outburst
within a supercycle was found to be faster than the later ones.
This suggests that the first normal outburst may most likely
be of the ``outside-in'' type while the later ones may be of the
``inside-out'' type. The occurrence of the ``inside-out''-type
outbursts in ER UMa may be understood as due to the mass
supply from the secondary to the inner part of the disk when
the disk tilted.

(4) The negative superhumps and the positive superhumps 
co-existed during the superoutburst although the signal of
the negative 
superhump was marginal in the first several days of the superoutburst.
By subtracting the signal of negative superhumps,
 we obtained the $O-C$ diagram for the positive superhumps
and by doing so we were able to
distinguish the stage A from the B ones.
A simple combination of
the positive and negative superhumps was not sufficient
to reproduce the complex profile variation.

(5) The number of normal outbursts in a supercycle of ER UMa
in 2011 and 2012 was found to be mostly
four, which  was smaller than that in other occasions
when negative superhumps were
not observed,  a result consistent with those reported for
ER UMa by \citet{zem13eruma} and for V1504 Cyg by \citet{osa13v1504cygKepler}. 
This can be understood as a result of the tilted disk:
when the disk is tilted, the gas stream coming from the secondary
arrives at the inner
region of the disk rather than at the disk edge and this reduces a frequent occurrence of the outside-in type normal outbursts.

(6) Some of the superoutbursts were triggered by
the normal outbursts ( i.e. precursor outbursts).
Positive superhumps started to grow during the declining
part of the precursor in such superoutbursts. 
 The phases of maxima were
found to be continuous
when a transition from the negative superhump to
the positive superhump occurred.

(6) We succeeded in detecting stage A superhumps
for the first time in ER UMa-type dwarf novae
during the precursor parts of three superoutburst.
Using the period of stage A superhumps,
we estimated $q$ of ER UMa to be 0.100(15).  The estimated 
$q$ and the known orbital period imply that ER UMa
 itself is in the standard evolutionary track of
cataclysmic variable stars. However, the mass transfer
late, $\dot{M}$ of ER UMa,  inferred from its short supercycle
length and frequent occurrence of normal outbursts, is much
higher than that of other SU UMa-type dwarf novae with similar
orbital periods.

 The authors are grateful to observers
of VSNET Collaboration and VSOLJ observers, who 
contributed observations,
covering long-term variations of ER UMa as long as 
two years.

   This work was partly supported by the Grant-in-Aid
``Initiative for High-Dimensional Data-Driven Science
through Deepening of Sparse Modeling'' from the Ministry
of Education, Culture, Sports, Science and Technology
(MEXT) of Japan, and Grants-in-Aid for the Global COE
Programme ``The Next Generation of Physics, Spun from
Universality and Emergence'' from the Ministry of Education,
Culture, Sports, Science and Technology of Japan.
We are very grateful to the referee of this paper, Osaki, Y.

\appendix

\section{The list of all observations}

\begin{longtable}{cccccc}
\caption{Log of observation list}\label{tab:log}
\hline
Start\commenta & End\commenta & $N$\commentd & obs\commente & sys\commentf & exp \\
\hline
\endhead
\hline
  \multicolumn{5}{l}{\commenta BJD$-$2455500.} \\
  \multicolumn{5}{l}{\commentd Number of observations.} \\
  \multicolumn{5}{l}{\commente Observer's code.} \\
  \multicolumn{5}{l}{\commentf Filter.  ``C'' means no filter (clear).} \\
\hline
\endfoot
\hline
  \multicolumn{5}{l}{\commenta BJD$-$2455500.} \\ %
  \multicolumn{5}{l}{\commentd Number of observations.} \\
  \multicolumn{5}{l}{\commente Observer's code.} \\
  \multicolumn{5}{l}{\commentf Filter.  ``C'' means no filter (clear).} \\
\hline
\endlastfoot
81.4612 & 81.7636 & 382 & deM & C & 60s \\ 
83.0675 & 83.3887 & 410 & Aka & C & 60s \\ 
83.9591 & 84.0043 & 100 & Kra & - & 30s \\ 
84.0048 & 84.0497 & 99 & Kra & C & 30s \\ 
84.0048 & 84.3783 & 888 & Ioh & C & 15s \\ 
84.6368 & 84.7775 & 306 & Kra & C & 30s \\ 
85.6340 & 86.0532 & 917 & Kra & C & 30s \\ 
86.0664 & 86.3767 & 421 & Aka & C & 60s \\
\end{longtable}

\section{The maximum timings of negative superhumps}

\begin{longtable}{cccc}\label{tab:nsh1}
\caption{Log of the maximum timings of negative superhumps supercycle 2011 S1}
\hline
E & The timings of maximum\commenta & $O-C$ & error  \\
\hline
\endhead
\hline
  \multicolumn{4}{l}{\commenta BJD$-$2455600} \\
\hline
\endfoot
\hline
  \multicolumn{4}{l}{\commenta BJD$-$2455600.} \\
\hline
\endlastfoot
0 & 91.0210 & 0.0010 & 0.0010 \\
1 & 91.0857 & 0.0034 & 0.0014 \\
2 & 91.1366 & $-$0.0081 & 0.0010 \\
3 & 91.2149 & 0.0079 & 0.0004 \\
4 & 91.2706 & 0.0012 & 0.0003 \\
5 & 91.3319 & 0.0002 & 0.0003 \\
8 & 91.5213 & 0.0026 & 0.0006 \\
9 & 91.5830 & 0.0020 & 0.0003 \\
\end{longtable}


\begin{longtable}{cccc}\label{tab:nsh2}
\caption{Log of the maximum timings of negative superhumps during supercycle 2011 S2}
\hline
E & The timings of maximum\commenta & $O-C$ & error  \\
\hline
\endhead
\hline
  \multicolumn{4}{l}{\commenta BJD$-$2455600} \\
\hline
\endfoot
\hline
  \multicolumn{4}{l}{\commenta BJD$-$2455600.} \\
\hline
\endlastfoot
0 & 142.3454 & $-$0.0006 & 0.0005 \\
1 & 142.4068 & $-$0.0016 & 0.0004 \\
2 & 142.4684 & $-$0.0023 & 0.0009 \\
3 & 142.5290 & $-$0.0040 & 0.0007 \\
10 & 142.9677 & $-$0.0017 & 0.0012 \\
11 & 143.0307 & $-$0.0011 & 0.0011 \\
13 & 143.1522 & $-$0.0042 & 0.0019 \\
14 & 143.2161 & $-$0.0026 & 0.0010 \\
\end{longtable}


\begin{longtable}{cccc}\label{tab:nsh3}
\caption{Log of the maximum timings of negative superhumps during supercycle 2011 S3}
\hline
E & The timings of maximum\commenta & $O-C$ & error  \\
\hline
\endhead
\hline
  \multicolumn{4}{l}{\commenta BJD$-$2455600} \\
\hline
\endfoot
\hline
  \multicolumn{4}{l}{\commenta BJD$-$2455600.} \\
\hline
\endlastfoot
0 & 191.4019 & $-$0.0031 & 0.0007 \\
1 & 191.4632 & $-$0.0041 & 0.0008 \\
2 & 191.5268 & $-$0.0028 & 0.0010 \\
14 & 192.2708 & $-$0.0065 & 0.0049 \\
15 & 192.3389 & $-$0.0007 & 0.0006 \\
16 & 192.4021 & 0.0003 & 0.0008 \\
17 & 192.4646 & 0.0004 & 0.0006 \\
31 & 193.3354 & $-$0.0010 & 0.0004 \\
\end{longtable}


\begin{longtable}{cccc}\label{tab:nsh4}
\caption{Log of the maximum timings of negative superhumps during 2012 S1}
\hline
E & The timings of maximum\commenta & $O-C$ & error  \\
\hline
\endhead
\hline
  \multicolumn{4}{l}{\commenta BJD$-$2455900} \\
\hline
\endfoot
\hline
  \multicolumn{4}{l}{\commenta BJD$-$2455900.} \\
\hline
\endlastfoot
0 & 0.6919 & 0.0119 & 0.0008 \\
1 & 0.7502 & 0.0080 & 0.0009 \\
16 & 1.6849 & 0.0092 & 0.0009 \\
17 & 1.7517 & 0.0138 & 0.0004 \\
31 & 2.6249 & 0.0158 & 0.0010 \\
32 & 2.6840 & 0.0126 & 0.0006 \\
33 & 2.7472 & 0.0136 & 0.0010 \\
72 & 5.1751 & 0.0145 & 0.0016 \\
\end{longtable}


\begin{longtable}{cccc}\label{tab:nsh5}
\caption{Log of the maximum timings of negative superhumps during supercycle S2}
\hline
E & The timings of maximum\commenta & $O-C$ & error  \\
\hline
\endhead
\hline
  \multicolumn{4}{l}{\commenta BJD$-$2455900} \\
\hline
\endfoot
\hline
  \multicolumn{4}{l}{\commenta BJD$-$2455900.} \\
\hline
\endlastfoot
0 & 43.5569 & $-$0.0071 & 0.0004 \\
1 & 43.6257 & $-$0.0007 & 0.0007 \\
2 & 43.6850 & $-$0.0037 & 0.0003 \\
3 & 43.7526 & 0.0015 & 0.0004 \\
8 & 44.0618 & $-$0.0010 & 0.0008 \\
9 & 44.1197 & $-$0.0055 & 0.0004 \\
10 & 44.1835 & $-$0.0041 & 0.0008 \\
11 & 44.2654 & 0.0154 & 0.0017 \\
\end{longtable}

\begin{longtable}{cccc}\label{tab:nsh6}
\caption{Log of the maximum timings of negative superhumps during supercycle 2012 S3}
\hline
E & The timings of maximum\commenta & $O-C$ & error  \\
\hline
\endhead
\hline
  \multicolumn{4}{l}{\commenta BJD$-$2455900} \\
\hline
\endfoot
\hline
  \multicolumn{4}{l}{\commenta BJD$-$2455900.} \\
\hline
\endlastfoot
0 & 97.1285 & 0.0055 & 0.0007 \\
1 & 97.1998 & 0.0144 & 0.0011 \\
2 & 97.2467 & $-$0.0010 & 0.0010 \\
3 & 97.3101 & 0.0001 & 0.0005 \\
10 & 97.7459 & $-$0.0005 & 0.0019 \\
11 & 97.8009 & $-$0.0078 & 0.0006 \\
12 & 97.8725 & 0.0014 & 0.0007 \\
13 & 97.9324 & $-$0.0010 & 0.0004 \\
\end{longtable}


\end{document}